\documentclass[final,onefignum,onetabnum]{siamart171218}

\usepackage[utf8]{inputenc} 

\usepackage{graphicx} 

\usepackage{CJKutf8}
\usepackage{booktabs} 
\usepackage{array} 
\usepackage{paralist} 
\usepackage{verbatim} 
\usepackage{subfig} 

\usepackage{epstopdf}
\usepackage{blkarray}
\usepackage{multirow}
\usepackage{amsmath}
\usepackage{harpoon}
\usepackage{amssymb}
\usepackage{mathrsfs}
\usepackage{graphics}
\usepackage{geometry}
\usepackage{epsfig}
\usepackage{mathtools}
\usepackage{comment}
\usepackage[final]{changes}
\usepackage{soul}
\usepackage[normalem]{ulem}

\def\dd{\mbox{d}}

\def\Dt{\Delta t}

\def\n{{\bf n}}

\def\u{{\bf u}}

\def\f{\frac}

\def\cb{\color{blue}}

\usepackage{wrapfig}
\usepackage{dsfont}
\usepackage{enumerate}

\usepackage{xcolor}

\begin{document}
\allowdisplaybreaks


\title{PDE models of adder mechanisms in cellular proliferation}
\author{Mingtao Xia\thanks{Department of Mathematics, UCLA, Los Angeles, CA, USA} \and
Chris D. Greenman\thanks{School of Computing Sciences, University of East Anglia, Norwich, UK}
\and Tom Chou\thanks{Depts. of Computational Medicine and Mathematics, UCLA, Los Angeles, CA, USA}}

\maketitle
\numberwithin{equation}{section}

\begin{abstract}
Cell division is a process that involves many biochemical steps and
complex biophysical mechanisms.  To simplify the understanding of what
triggers cell division, three basic models that subsume more
microscopic cellular processes associated with cell division have been
proposed. Cells can divide based on the time elapsed since their
birth, their size, and/or the volume added since their birth -- the
timer, sizer, and adder models, respectively. Here, we propose unified
adder-sizer models and investigate some of the properties of different
adder processes arising in cellular proliferation.  Although the
adder-sizer model provides a direct way to model cell population
structure, we illustrate how it is mathematically related to the
well-known model in which cell division depends on age and size.
Existence and uniqueness of weak solutions to our 2+1-dimensional PDE
model are proved, leading to the convergence of the discretized
numerical solutions and allowing us to numerically compute the
dynamics of cell population densities. We then generalize our PDE
model to incorporate recent experimental findings of a system
exhibiting mother-daughter correlations in cellular growth rates.
Numerical experiments illustrating possible average cell volume blowup
and the dynamical behavior of cell populations with mother-daughter
correlated growth rates are carried out. Finally, motivated by new
experimental findings, we extend our adder model cases where the
controlling variable is the added size between DNA replication
initiation points in the cell cycle.
  \end{abstract}



\section{Introduction}

How cells regulate and maintain their sizes, as well as sizes of their
appendages is a longstanding research topic in cell biology. Besides
growth of an individual cell, the size distributions within a
population of cells are also a quantity of interest.  When considering
proliferating cell populations, individual cell growth is interrupted
by cell division events that generate smaller daughter cells. The
biological mechanisms that control when and how a cell divides are
complex and involve many steps such as metabolism, gene expression,
protein production, DNA replication, chromosomal separation (for
eukaryotic cells), and fission or cell wall formation
\cite{MAALOE1973,NOTCH1996,CHANDLER_BROWN,CADART,OSKAR2017}.  These
processes are regulated and may involve intricate biochemical
signaling.

Despite the complexity of cell growth and the cell cycle, three simple
hypotheses for the underlying mechanisms of cell division have been
proposed.  Cell division can be governed by cell age $a$, cell volume
$x$ \cite{SINKO1967}, or added volume since birth $y$
\cite{ECOLI_MODEL,JUN2015}. The division mechanism employed by a type
of cell may be interrogated by tracking the volumes $x$, added volumes
$y$, and ages $a$ during division events. Volume growth of an
individual cell can be straightforwardly measured and can be modeled
by an effective empirical law such as $\dot{x} = g(a,x, y, t)$. A
commonly used approximation that is supported by observations is the
exponential growth law $g(x) = \lambda x$ \cite{SMK1958}.

To describe population-level distributions, PDE approaches have been
developed.  For example, the timer model, in which the cell division
rate depends only on age of the cell is described by the classic
McKendrick equation for $n(a,t)$, the expected density of cells at age
$a$ and time $t$ \cite{MCKENDRICK,VONFOERSTER}. The McKendrick
``transport'' equation for the cell density takes the form
$\partial_{t}n(a,t) + \partial_{a}n(a,t) = -(\mu(a)+\beta(a))n(a,t)$,
in which $\beta(a)$ and $\mu(a)$ are age-dependent birth and death
rates, respectively.  The associated boundary condition $n(t,0) =
2\int_{0}^{a}\beta(s)n(s,t)\dd s$ describes the birth of zero-age
cells. Fully demographically stochastic versions of the timer model
have also been recently developed
\cite{GREENMANPRE,GREENMANJSP,greenman2017path}.

The timer (or age-dependent) model does not explicitly track cell
sizes, but PDE models incorporating sizer mechanisms have been
developed \cite{PERTHAME2008,DOUMIC_INVERSE2009,DOUMIC2014}.  In these
studies size-dependent birth rates $\beta(x)$ are pertinent.
Depending on the form of $\beta(x)$, cells can diverge in size $x$ in
the absence of death \cite{DOUMIC_M3AS}.  Existence and uniqueness of
weak solutions to timer and sizer models have been proved for certain
boundary and initial conditions.  These types of structured population
equations can be partially solved using the method of characteristics
but the boundary conditions can only be reduced to a Volterra-type
integral equation \cite{PERTHAME2008,GREENMANJSP}.

Much like a general growth law $g(a,x,y,t)$ that can depend on age,
size, added size, and time, the three distinct mechanisms of cell
division need not be mutually exclusive. In this paper, we mainly
focus, at the cell population level, on the cell division mechanism
that incorporates the added volume, or the so-called the ``adder.''
This mechanism, in which the cell seems to use added size as the
factor controlling its division, has been indicated in many recent
experimental studies. Specifically, apart from the sizer and the timer
models, the adder mechanism has been recently shown to be consistent
with E. coli division \cite{MAALOE1973,JUN2015,ECOLI_MODEL} and can be
motivated by an initiator accumulation mechanism distinct from those
used to justify sizers or timers \cite{JUN2015,CHANDLER_BROWN}.

We will introduce the PDE model that describes cell population
structure under the adder mechanism, which we describe as the
``adder-sizer'' PDE model, and show its connection to the classical
``timer-sizer'' PDE model that involves cell age and size as
controlling parameters.  The proof of the existence and uniqueness of
a weak solution to the proposed three-dimensional ``adder-sizer'' PDE
turns out to be more complex than the proof for the timer and/or sizer
counterparts \cite{PERTHAME2008}. Our proof leads to the convergence
of the numerical solutions to the adder-sizer PDE, allowing us to
numerically evaluate the corresponding structured cell populations,
facilitating further analysis, exploration of possible ``blowup''
behavior, and generalizations of the model.  Stochastic Monte-Carlo
simulations of the corresponding stochastic process are also generated
and compared with numerical results for $n(x,y,t)$ and division-event
densities.

Next, we propose an extension to the adder-sizer model that
incorporates cellular growth rates that are correlated across
successive generations.  Changes in growth rates at the single-cell
level have been explored using stochastic mapping methods
\cite{DOUMIC_ESTIMATION2015, AMIR2017}. By numerically solving the
PDE, we found out that the population-averaged growth rate are larger
when correlations between mother and daughter cell growth rates are
larger.  Finally, we generalized the adder model to include a
different two-phase PDE system which could describe the latest
``initiation adder'' mechanism, which states that the added mechanism
takes effect on the cell's size at initiation instead of division in
\cite{JUN2019}. In contrast to the single-PDE division adder model, a
model describing the initiation adder mechanism requires two coupled
PDEs.

To model cell size control, stochastic maps that relate daughter cell
sizes to mother cell sizes have been developed
\cite{KESSLER2017,SINGH2017}. These models describe how cell sizes
evolve with generation and can interpolate among timer, sizer, and
adder mechanisms.  Kessler and Burov \cite{KESSLER2017} assumed
stochastic growth which lead to a stochastic map with multiplicative
noise. They found that an adder mechanism can admit ``blow-up'' in
which the expected cell sizes can increase without bound with
increasing generation observed experimentally in filamentous
bacter. Modi \textit{et al.} \cite{SINGH2017} assume additive noise
and do not find blow-up in an adder model.  Stochastic maps of
generational cell size do not describe population-level distributions
in size or age.

\section{Adder-sizer PDE models}

Here, we introduce adder-sizer PDE models and generalize them to
describe recently observed characteristics of population-level
bacterial cell division.  An adder-sizer model is one that
incorporates a cell division rate $\beta(x,y,t)$ and a single-cell
growth rate $g(x,y,t)$ that, instead of depending on a cell's age $a$,
are functions of cell size $x$ and a cell's volume \textit{added since
  birth} $y$. Such an adder-sizer PDE model can be developed by
defining $n(x,y, t)\dd x \dd y$ as the mean number of cells with size
in $[x,x+\dd x]$ and added volume in $[y, y+\dd y]$.  As cells have
finite size and their added volume must be less than total size,
$n(x\leq 0,y,t)=n(x,y\geq x,t)=0$. A derivation similar to that given
in \cite{Metz1986} for the sizer model yields a transport equation of
the form

\begin{equation}
\begin{array}{l}
\displaystyle {\partial n(x,y,t)\over \partial t} +
{\partial [g(x,y,t)n(x,y,t)]\over \partial x} + {\partial [g(x,y,t)n(x,y,t)]
\over \partial y}
= - \beta(x,y,t) n(x,y,t)
\label{AS_PDE}
\end{array}
\end{equation}
for the adder-sizer PDE. Here, we have neglected the effects of death,
which can be simply added to the right-hand-side of Eq.~\ref{AS_PDE}.

\begin{figure}[htb]
\begin{center}
\includegraphics[height=1.65in]{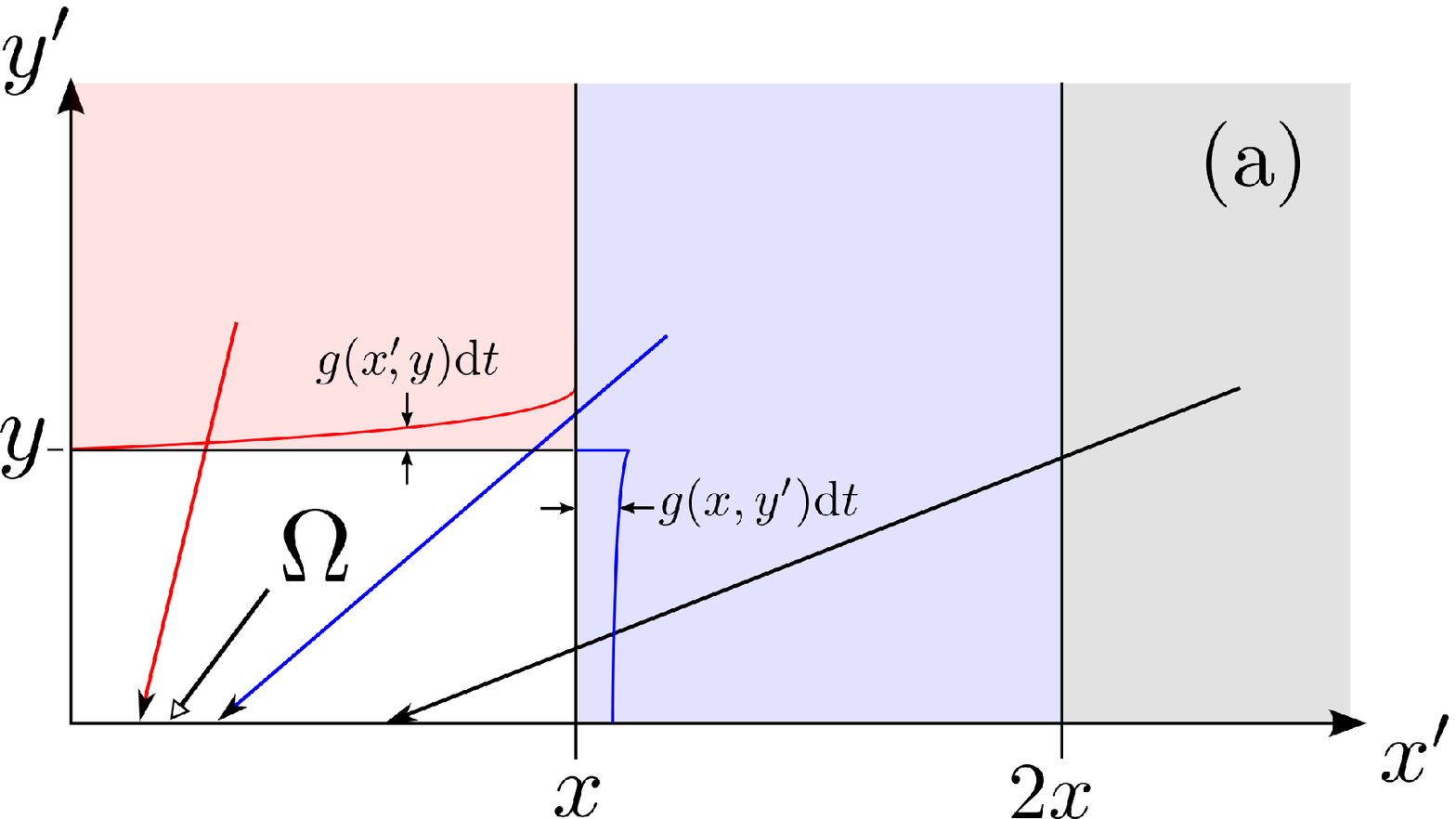}\hspace{9mm}\includegraphics[height=1.6in]{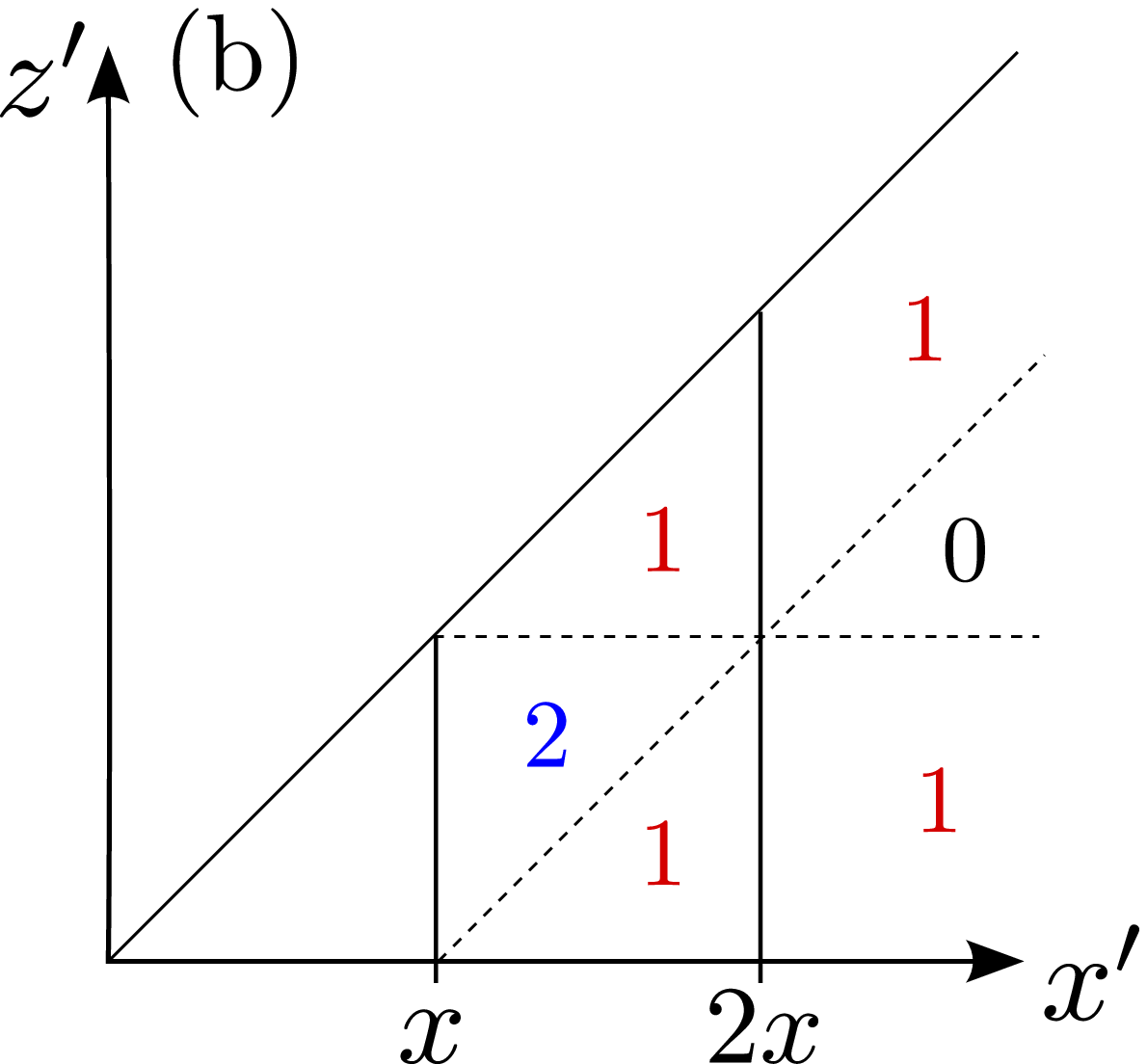}
\caption{The size and added-size state space for cell populations.
  The expected total number of cells at time $t$ with added size
  within $[0, y]$ and volume (or ``size'') within $[0, x]$ is defined
  as $N(x,y,t)$. Over an increment in time $\dd t$, the domain $\Omega
  = [0,y]\times [0,x]$ infinitesimally distorts $\Omega \to \Omega +
  \dd \Omega$ through the growth increment $g\dd t$.  The total
  population within this distorted domain changes only due to birth
  and death. Cells within $\Omega$ that divide always give rise to two
  daughters within $\Omega$, leading to a net change of $+1$ cell.
  (b) The $z'$ and $x'$ domains of the differential birth rate
  function $\tilde{\beta}(x',y',z',t)$.  Cells outside of $\Omega$ can
  contribute a net +1 or +2 cells in $\Omega$ depending on the
  division patterns defined in the depicted regions.}
\label{A_X}
\end{center}
\end{figure}

To explicitly outline our general derivation, consider the total
population flux into and out of the size and added size domain
$\Omega$ shown in Fig.~\ref{A_X}(a) and define
$\tilde{\beta}(x',y',z',t) \dd z'$ as the rate of fission of cells of
size $x'$ and added size $y'$ to divide into two cells, one with size
in $[z', z'+\dd z']$ and the other with size within $[x'-z',
  x'-(z'+\dd z')]$.  For binary fission, conservation of daughter cell
volumes requires $\tilde{\beta}(x',y',z',t) \equiv
\tilde{\beta}(x',y', x'-z',t)$.  This differential division function
allows mother cells to divide into two daughter cells of differing
sizes (asymmetric division), a process that has been observed in
numerous contexts \cite{HORVITZ,NOTCH1996,KNOBLICH}. We also assume
that daughter cells must have positive size so
$\tilde{\beta}(x',y',z'=0,t) = \tilde{\beta}(x',y',z'=x',t)=0$.

The change in the number of cells in $\Omega$ due to fission can arise
in a number of ways. First, if a cell in $\Omega$ divides, it can only
produce two cells with size less than $x$. Thus, such fission events
lead to a net change of $+1$ in the number of cells with $y=0$ and
size in $[0,x]$. 
%
%
If a cell with size within $[0,x]$ but with added size $>y$ divides, it
creates two cells with added size $y=0$ and size within $[0,x]$, leading to a
net change of $+2$ cells. 

For cells with any added size $y'>0$ but with size $x'>x$, we have two
subcases. If the dividing cell has size $x < x' < 2x$, it will produce
one daughter cell in $\Omega$ if a daughter cell has size $0< z' <
x'-x$ or $x < z' <x'$ as shown in Fig.~\ref{A_X}(b). If $x'-x < z' <
x$, both daughter cells have size $<x$.  Finally, if the dividing cell
has size $x' > 2x$, at most one daughter will have size $x'<x$ (see
Fig.~\ref{A_X}(b)).  Upon simplifying the above birth terms by using
$\int_{0}^{x'}\!\dd z' = \int_{0}^{x}\!\dd z'+ \int_{x}^{x'}\!\dd z'$
for $x'>x$ and the symmetry $\tilde{\beta}(x',y',z',t) =
\tilde{\beta}(x',y',x'-z',t)$, we combine terms to balance
proliferation with transport and find

\begin{align}
\int_{0}^{x}\!\dd x' \int_{0}^{y}\!\dd y'
\,{\partial n(x',y',t)\over \partial t}
+ & \int_{0}^{x}\!\dd x'\, g(x',y,t)n(x',y,t)  + \int_{0}^{y}\!\dd y'\,
g(x,y',t) n(x,y',t) \nonumber \\
\:  & \hspace{1cm} = \int_{0}^{\infty}\!\dd y' 
\int_{0}^{x}\!\dd x' \int_{0}^{x'}\!\dd z'
\,\tilde{\beta}(x',y'z',t)n(x',y't) \nonumber \\
\: &  \hspace{1.4cm} + \int_{y}^{\infty}\!\dd y' \int_{0}^{x}\!\dd x'
\int_{0}^{x'}\!\dd z'\,\tilde{\beta}(x',y,z',t)n(x',y',t) \nonumber \\
\: &  \hspace{1.4cm} + 2\int_{0}^{\infty}\!\dd y' \int_{x}^{\infty}\!\dd x'
\int_{0}^{x}\!\dd z'\,\tilde{\beta}(x',y',z',t)n(x',y',t).
\label{DNDT2}
\end{align}
Upon taking the derivatives ${\partial^{2} \over \partial x \partial
  y}$, we find the PDE given in Eq.~\ref{AS_PDE} where the total
division rate is defined by $\beta(x,y,t) \coloneqq
\int_{0}^{x}\tilde{\beta}(x,y,z,t)\dd z$. For the boundary condition
at $y=0$, we take the derivative $\partial/\partial x$ and set $y\to
0^{+}$ to find

\begin{equation}
g(x,y=0,t)n(x,y=0,t) = 2\int_{x}^{\infty}\!\!\dd x'
\int_{0}^{x'}\!\!\dd y'
\,\tilde{\beta}(x',y',z=x,t)n(x',y',t).
\label{AS_BC}
\end{equation}
The other boundary condition defined by construction is
$n(x,x, t) = 0$.

In the special restricted case of symmetric cell division,
$\tilde{\beta}(x,y,z,t) = \beta(x,y,t)\delta(z - x/2)$,
and boundary condition of the adder-sizer model reduces to 

\begin{equation}
g(x,y=0,t)n(x, y=0,t)= 4\int_0^{2x}\beta(2x, y', t)n(2x, y', t) \dd y'.
\end{equation}
The above derivation provides an explicit boundary condition
representing newly born cells that may be asymmetric in birth size.
Quantities such as the total cell population $N(t)$ and the mean total
biomass $M(t)$ (the total volume over all cells) can be easily
constructed from the density $n(x,y,t)$:

\begin{equation}
\displaystyle
N(t)  =\int_0^{\infty}\!\dd x\int_{0}^{x}\!\!\dd y\, n(x, y,t), \quad
M(t)  =\int_0^{\infty}\!\dd x \int_{0}^{x}\!\!\dd y\, x n(x, y, t).
\label{NM}
\end{equation}
Higher moments of the total volume can also be analogously defined.  By
applying these operations to Eq.~\ref{AS_PDE} and using the boundary
condition (Eq.~\ref{AS_BC}), we find the dynamics of the total
population and biomass

\begin{equation}
\frac{\dd N(t)}{\dd t} = \int_0^{\infty}\dd x \int_{0}^{x}\!\!\dd y\,
\beta(x, y, t)n(x, y, t),\,\,  \frac{\dd M(t)}{\dd t} = 
\int_{0}^{\infty}\!\!\dd x \int_{0}^{x}\!\!\dd y\,
g(x,y,t)n(x, y, t).
\label{MNdot}
\end{equation}

Finally, we also define the distribution of division events over the
size and added size variables, accumulated over a time $T$:

\begin{equation}
  \rho_{\rm d}(x,y,T) \displaystyle = \frac{\displaystyle
    \int_{0}^{T}\beta(x,y,t)n(x,y,t)\dd t}{
\displaystyle \int_{0}^{T}\!\!\dd t \int_{0}^{\infty}\!\!\dd x'\int_{0}^{x}\!\!\dd y'\,
\beta(x',y',t)n(x',y',t)}.
\label{RHOD}
\end{equation}

\subsection{Division probability and connection to time-sizer model} 

In general, the birth rate functions $\tilde{\beta}(x,y,z,t)$ and
$\beta(x,y,t)$ associated with adder-sizer models can take many forms
that make biological sense. However, some classes of $\beta(x,y,t)$
may allow the adder-sizer model to be transformed into the well-known
``sizer-timer'' structured population model \cite{SINKO1967}.  To
illustrate the relationship, we consider a division rate function
$\beta$ which depends explicitly only on age $a$ and see how it could
be converted to a function of size and added size.

For a cell born at time $t_{0}$, the probability that the cell splits
within time $[a,a+\dd a]$ is defined by $\gamma(a; \bar{a})\dd a$. In
the absence of death, to ensure that any single cell will eventually
split, $\int_{0}^{\infty}\gamma(a; \bar{a})\dd a = 1$.  Reasonable
choices for $\gamma(a; \bar{a})$ are Gamma, lognormal, or normal
distributions. Without loss of generality, we propose a simple gamma
distribution for $\gamma(a; \bar{a})$:

\begin{equation}
\gamma(a;\bar{a}) =
\frac{1}{a\Gamma((\bar{a}/\sigma_{a})^{2})}\exp\left[-{a\bar{a}\over
    \sigma_{a}^{2}}+\left({\bar{a}\over \sigma_{a}}\right)^{2}\ln
  \left({a\bar{a}\over \sigma_{a}^{2}}\right)\right],
\label{GAMMAA}
\end{equation}
where $\bar{a}$ is the mean division age and $\sigma_{a}^{2}$ is the
variance.  This type of distribution can be derived from the sum of
independent, exponentially distributed ages.

For determinisitic exponential growth $g=\lambda{x}$, age $a$ and the
parameter $\bar{a}$ can be explicitly expressed in terms of $x,y$ and
possibly other fixed parameters:
\begin{equation}
  a(x,y) = \frac{1}{\lambda}\ln\left({x\over x-y}\right), \quad 
\bar{a}(x, y) = \f{1}{\lambda}\ln\left(\frac{x-y+\Delta}{x-y}\right),
  \label{A_XY}
\end{equation}
in which $\Delta$ is the fixed added size parameter that represents
the adder mechanism.

With $a(x,y)$ and $\bar{a}(x,y)$ defined in Eqs.~\ref{A_XY}, the
division rate function $\beta(x,y)$ can be expressed in terms of $x$
and $y$ by using the splitting probability
$\gamma(a(x,y);\bar{a}(x,y))$:

\begin{equation}
  \beta(x,y,t) = \frac{\gamma(a(x,y); \bar{a}(x, y))}{1-\int_{0}^{a(x,y)}\!\dd a' \gamma(a'; \bar{a}(x, y))}.
  \label{BETA_GAMMA}
\end{equation}

\begin{figure}[h!]
\begin{center}
\includegraphics[height=2.2in]{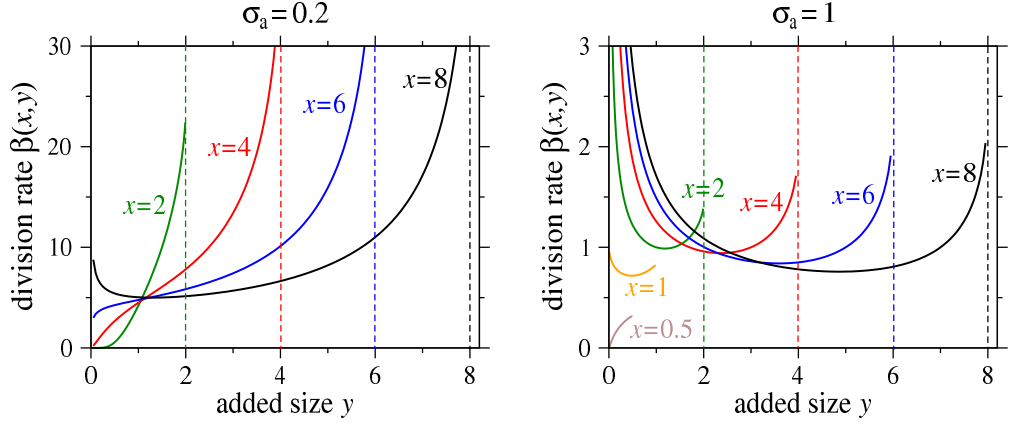}
\caption{The size and added-size dependent rate $\beta(x,y)$
  constructed using a gamma distribution for the splitting probability
  $\gamma$ (Eq.~\ref{GAMMAA}) and Eq.~\ref{BETA_GAMMA}. We show
  projections at fixed values of $x$. In (a) the parameters are
  $\sigma_{\rm a} = 0.2$, while in (b) $\sigma_{\rm a}=1$. Note the
  difference in scale and that $\gamma(a)$ with a higher standard
  deviation leads to a lower overall cell division rate $\beta$. When
  $x$ is large, $\bar{a}$ defined in \ref{A_XY} is small, a nonzero
  division rate $\beta(x,y\to 0) > 0$ arises indicating that large
  newborn cells divide quickly to control size across the
  population. This particular feature arises from our construction of
  $\beta$ as a hazard function. Modifying birth rate at small values
  of $y$ so that $\beta(x, y= 0) \to 0$ will not qualitatively change
  the predicted densities as long as the birth rate peak persists at
  small $y$.}
\label{BETA_PLOT}
\end{center}
\end{figure}

Assuming this ``hazard function'' form of a growth law, cells born at
small initial size $x(0) = x_{0} = x-y$ take longer time to divide,
while cells born with large size split sooner. Using the gamma
distribution, we find a division rate of the form

\begin{equation}
\beta(x,y) = {\Gamma\left({\bar{a}^{2}(x,y)\over
    \sigma_{a}^{2}}\right)\gamma(a(x,y); \bar{a}(x, y))\over
\Gamma\left({\bar{a}^{2}(x,y)\over \sigma_{a}^{2}},
           {a(x,y)\bar{a}(x,y)\over \sigma_{a}^{2}}\right)},
\end{equation}
where $\Gamma(\cdot, \cdot)$ is the upper incomplete gamma function.
We plot two examples of the time-independent rate $\beta(x,y)$ in
Fig.~\ref{BETA_PLOT}.

With $\beta(x,y,t)$ defined, we still need to construct the full
fission rate $\tilde{\beta}$, which we will assume is a product of the
overall division rate $\beta(x,y,t)$ and a differential division
probability.  The simplest model is to assume that the differential
division probability $h(r)$ is a function of only the ratio $r$
between the size of the daughter cell and that of the mother cell, and
independent of the cell size just before division. Thus,

\begin{equation}
\tilde{\beta}(x,y,z,t) = \beta(x,y,t) h(z/x)/x,
\label{TildeBeta}
\end{equation}
where $r \equiv z/x \in[0,1]$.  The boundary condition
(Eq.~\ref{AS_BC}) can thus be written in the form

\begin{equation}
  g(x,0,t)n(x,0,t) =2 \int_{x}^{\infty}\!\!\dd x' \int_{0}^{1}\!\!\dd
s\,\beta(x',sx',t)h(x/x') n(x',sx',t).
\end{equation}
A reasonable model for $h(r=x/x')$ is a lognormal form that is symmetric about
$r=1/2$:

\begin{equation}
h(r) \displaystyle = {h_{0}(r) + h_{0}(1-r)\over Z(\sigma_r,\delta)}, \quad 
h_{0}(r) \displaystyle = e^{-\frac{(-\delta + \ln
    r)^{2}}{2\sigma_{r}^{2}}}e^{-\frac{\ln^{2}(1-r)}{2\sigma_{r}^{2}}},
\label{HR}
\end{equation}
where the parameters $\delta$ and $\sigma_{r}$ determine the bias and spread of
the daughter cell size distribution,  and the normalization constant is
$Z(\sigma_{r},\delta)= \int_{0}^{1} (h_{0}(r)+h_{0}(1-r)) \dd r$.

\subsection{Numerical Implementation and Monte-Carlo Simulations}
With the differential birth rate function $\tilde{\beta}$ defined, we
can now consider the implementation of numerical solutions to
Eqs.~\ref{AS_PDE} and \ref{AS_BC} as well as event-based simulations
of the underlying corresponding stochastic process.  Since a typical
initial condition may not be smooth, a classical solution to
Eqs.~\ref{AS_PDE} and \ref{AS_BC} may not exist.  Thus, we provide a
proof of existence and uniqueness of the weak solution to
Eqs.~\ref{AS_PDE} and \ref{AS_BC} in Appendix \ref{A}.  We show
convergence of a discrete approximation to our problem, allowing us to
confidently numerically approximate the weak solution.

The numerical approximation to the weak solution will be based on an
upwind finite difference scheme in which both $x$ and $y$ are
discretized with step size $h$.  We define locally averaged functions
by

\begin{equation}
f_{i+\frac{1}{2},j+\frac{1}{2}} \coloneqq 
\frac{1}{h^{2}}\int_{ih}^{(i+1)h}\!\!\dd x \int_{jh}^{(j+1)h}\!\!\dd y\, f(x,y,t),
\end{equation}
where $f(x,y,t)$ can represent $n(x,y,t)$, $g(x,y,t)$, or
$\beta(x,y,t)$.  Similarly,

\begin{equation}
\tilde{\beta}_{i+\frac{1}{2},j+\frac{1}{2}}((s+\frac{1}{2})h, t) 
= h^{-3}\int_{ih}^{(i+1)h}\!\!\dd x
\int_{jh}^{(j+1)h}\!\!\dd y \int_{kh}^{(k+1)h}\!\!\dd z\,
\tilde{\beta}(x,y,z,t)
\end{equation}
in the domain $i, j \geq 0$ and $j,k < i$.  The discretization of the
transport equation can be expressed as

\begin{equation}
\begin{array}{l}
\frac{n_{i+\frac{1}{2},j+\frac{1}{2}}(t+\Dt) - 
n_{i+\frac{1}{2},j+\frac{1}{2}}(t)}{\Dt} + 
\frac{g_{i+1,j+\frac{1}{2}}\tilde{n}_{i+1,j+\frac{1}{2}}-g_{i,j}\tilde{n}_{i,j+\frac{1}{2}}}{h}
+ \frac{g_{i+\frac{1}{2},j+1}\tilde{n}_{i+\frac{1}{2},j+1}
-g_{i+\frac{1}{2},j}\tilde{n}_{i+\frac{1}{2},j}}{h} \noindent \\
\:\hspace{3cm}= -\beta_{i+\frac{1}{2},j+\frac{1}{2}} n_{i+\frac{1}{2},j+\frac{1}{2}}(t),
\label{NumSch}
\end{array}
\end{equation}
for $1\leq i,j \leq L$, where $Lh$ is the maximum size which we take
sufficiently large such that $n_{i,j > K}(t=0) = 0, n_{i\leq
  j}=0$. We also set $g_{i+\frac{1}{2},i}=0$ to prevent density
  flux out of the $y<x$ domain.  In Eq.~\ref{NumSch}, $g_{i+1,
  j+\frac{1}{2}}(t)$ can be taken as $g((i+1)h, (j+\frac{1}{2})h, t)$
while $\tilde{n}_{i+1, j+\frac{1}{2}}(t) = \int_{jh}^{(j+1)h}\!\!\!
\dd{y}\, n((i+\frac{1}{2})h, y, t)$ is a finite-volume numerical
approximation to $\int_{jh}^{(j+1)h}\!\!\!  \dd{y}\, n((i+1)h, y,
t)$. The discretized version of the boundary condition
(Eq.~\ref{AS_BC}) can be expressed as

\begin{equation}
g_{i+\frac{1}{2},0}n_{i+\frac{1}{2},0}(t)
=2h^{2}\sum_{k=i+1}^{L}\sum_{j=0}^{k-1}
\tilde{\beta}_{k+\frac{1}{2},j+\frac{1}{2}}((i+\frac{1}{2})h, t)
n_{k+\frac{1}{2},j+\frac{1}{2}}(t).
\label{NumSch2}
\end{equation}
The full explicit discretization scheme for the numerical calculation
is provided in Appendix B.

Direct Monte-Carlo simulations of the birth process are also performed
and compared with our numerically computed deterministic distributions
(see Appendix C). We construct a list of cells and their associated
sizes and their sizes at birth.  This list is updated at every time
step $\Dt$. The cell sizes grow according to $g(x,y,t)$. If a cell
divides, the initial sizes of the daughter cells are randomly chosen
according to the distribution $h(z/x)$.  The daughter cells then
replace the mother cell in the list. Simulations of the underlying
stochastic process results in, at any given time, a collection of
cells, each with a specific size and added size. This collection of
cells represents a realization of the population that should be
approximated by the distributions that are solutions to
Eqs.~\ref{AS_PDE} and \ref{AS_BC}.

\section{Analysis and Extensions}

In this section, we numerically investigate the adder-sizer model and
plot various cell population densities and birth event distributions
under different parameter regimes. We also show the consistency of
numerical solutions of the adder-sizer PDE with results from direct
Monte-Carlo simulations of the corresponding stochastic process, which
demonstrates that numerical solutions of the linear PDE model for cell
population is in agreement with single-cell level stochastic models.
After investigating birth rate parameters that can lead to blow-up of
population-averaged cell sizes, we extend the basic adder model to
include mother-daughter growth rate correlations and processes that
measure added size from different points in the cell cycle,
\textit{i.e.}, an initation-adder model.

\subsection{Cell and division event densities}

\begin{figure}[h!]
\begin{center}
\includegraphics[height=3.75in]{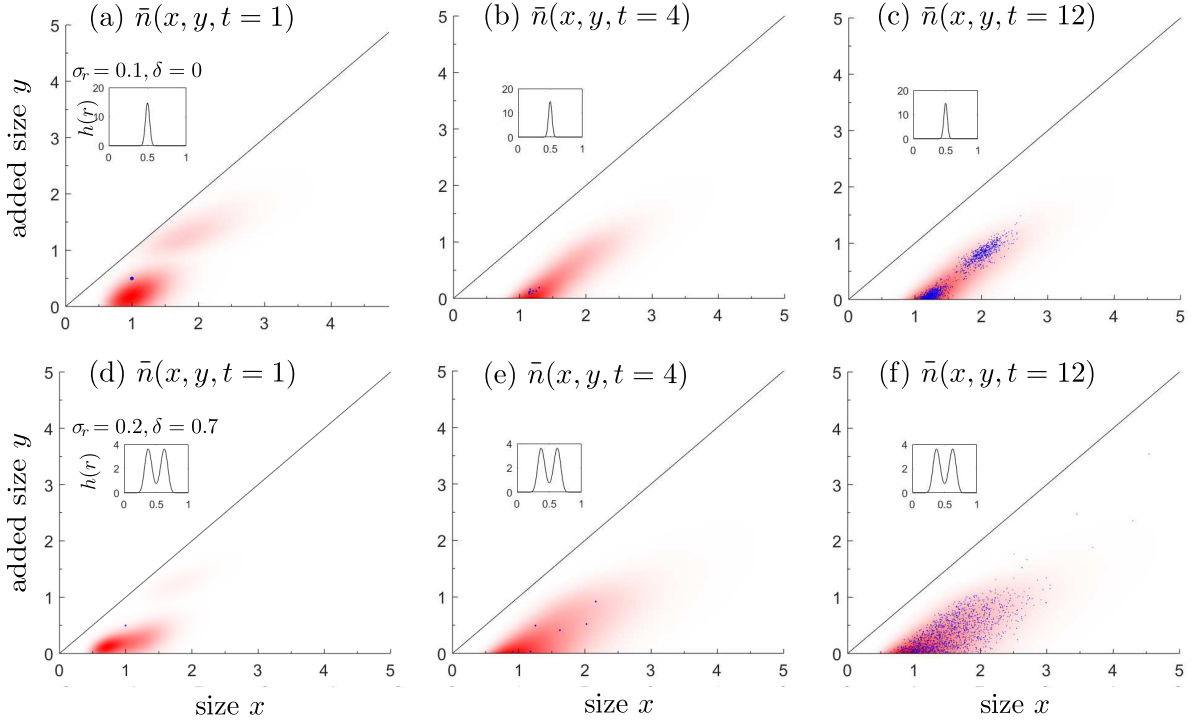}
\caption{Numerically computed densities $\bar{n}(x,y,t)=
  n(x,y,t)/N(t)$ using $g(x,y,t) = \lambda x$ and
  $\tilde{\beta}(x,y,z,t)$ defined by Eqs.~\ref{BETA_GAMMA},
  \ref{GAMMAA}, and \ref{HR}. For all plots, we use $\sigma_{a} = 0.1$
  in $\gamma(a)$ (Eq.~\ref{GAMMAA}) and rescale size in units
  of $\Delta$.  In (a-c), we use the sharp, single-peaked differential
  division function $h(r)$ shown in the inset ($\sigma_{r}=0.1,
  \delta=0$) and plot $\bar{n}(x,y,1), \bar{n}(x,y,4)$, and
  $\bar{n}(x,y,12)$, respectively. In (d-f), we plot the densities
  using a broad (in fact, double-peaked) differential division
  function $h(r)$ with parameters $\sigma_{r}=0.2, \delta=0.7$. In all
  calculations, we assumed an initial condition corresponding to a
  single newly born ($y=0$) cell with size $x=1$. For more asymmetric
  cell division in (d-f), the density spreads faster. In these cases,
  the densities closely approach a steady-state distribution by about
  $t=12$.  Also shown in each plot are realizations of Monte-Carlo
  simulations of the discrete process. Individual cells are
  represented by blue dots which accurately sample the normalized
  continuous densities $\bar{n}(x,y,t)$.}
\label{DENSITY}
\end{center}
\end{figure}

\begin{figure}[h!]
\begin{center}
\includegraphics[height=3.75in]{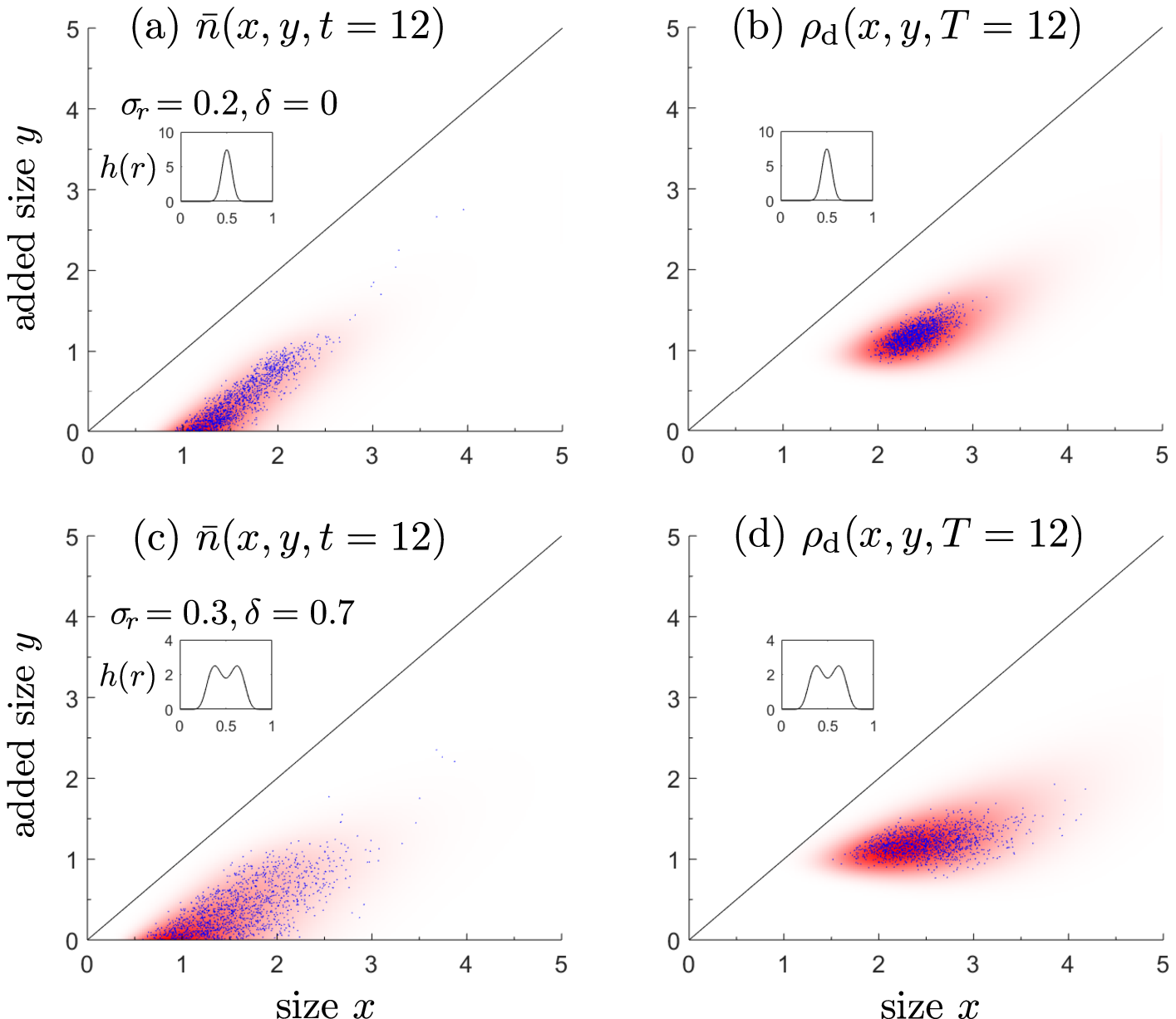}
\caption{Comparison of cell densities $\bar{n}(x,y,t)$ and cell
  division event densities $\rho_{\rm d}(x,y,T)$ (Eq.~\ref{RHOD}). The
  standard deviation $\sigma_{a} = 0.1$ is used in all calculations. In (a) and
  (b) we plot $\bar{n}(x,y,t=12)$ and $\rho_{\rm d}(x,y,T)$ using
  $\sigma_{r}=0.2, \delta=0$ while in (c) an (d) we used a broader
  differential division function in which $\sigma_{r}=0.3,
  \delta=0.7$. Realizations from Monte-Carlo simulations are
  overlayed. In (b) and (d), divisions are accumulated up to time
  $T=12$.}
\label{EVENTDENSITY}
\end{center}
\end{figure}

We evaluated our adder-sizer PDE model by using the division rate
given in Eq.~\ref{BETA_GAMMA} and first assuming the simple and
well-accepted growth function $g(x,y,t) = \lambda x$. Fig.~\ref{DENSITY}
shows the numerical results for the density $\bar{n}(x,y,t) =
n(x,y,t)/N(t)$ at successive times $t=1,4,12$, respectively.
Stochastic simulations of the underlying process yield cells
populations consistent with the deterministic densities derived from
the PDE model.  In Fig.~\ref{EVENTDENSITY}, we compare the cell
densities $\bar{n}(x,y,t)$ the division event densities $\rho_{\rm
  d}(x,y,T)$ for two different differential division functions
$h(r)$. As before, the more asymmetric the division the broader the
cell and event densities.

\subsection{Cell Volume Explosion}

At the single-cell level, a stochastic map model by Kessler and Burov
assumed a multiplicative noise and predicted that cell sizes can
eventually grow without bound, in agreement with what was
experimentally observed for filamentous bacteria \cite{KESSLER2017}.
However, stochastic maps of generational cell size do not capture
population-level distributions in size or age. In this subsection, we
will numerically explore how a possible "blowup" in the
population-averaged cell volumes.  Within PDE models that describe
population distributions, timer and sizer mechanisms have been shown
to exhibit blow-up depending on properties of the birth rate
$\beta(a,x)$
\cite{BERNARD2016,DOUMIC_ESTIMATION2015,DOUMIC_M3AS}. Analysis of the
conditions on full differential division rate $\tilde{\beta}(x,y,z,t)$
that would result in blow-up in the ``adder-sizer'' PDE model is more
involved. Here, we provide only a heuristic argument for sufficient
conditions for blow-up.

First, we characterize the shape of the densities in the adder-sizer
model.  In the analogous McKendrick equation \cite{IANNELLI1995} one
can investigate the age profile defined by dividing the number density
by the total population size.  The long term age profile may be stable
even when the total population size continuously increases.  We take a
similar approach here by analyzing $\bar{n}(x,y,t) = n(x,y,t)/N(t)$
where $N(t)$ is given by Eq.~\ref{NM}. Writing the adder-sizer PDE in
terms of $\bar{n}$, we find


\begin{equation}
\frac{\partial \bar{n}}{\partial t} + \frac{\bar{n}}{N}\frac{\dd N}{\dd t} + 
\frac{\partial (g\bar{n})}{\partial x}+\frac{\partial(g\bar{n})}{\partial y} 
= -\beta\bar{n}.
\label{NBAR}
\end{equation}
Integrating this equation over $x,y$ leads to $\dot{N}/N =
\int_{0}^{\infty}\!\!\dd x \int_{0}^{x}\!\!\dd y \, \beta\bar{n}$,
which can be substituted into the first term in Eq.~\ref{NBAR} to
yield the nonlinear PDE

\begin{equation}
\frac{\partial \bar{n}}{\partial t} + {\partial (g\bar{n})\over \partial x}
+ {\partial (g\bar{n})\over \partial y} = -\left(\beta + 
\int_{\Omega}\beta \bar{n}\right)\bar{n}.
\label{NBARNONLINEAR}
\end{equation}

A number of standard approaches may be applied to analyze
Eq.~\ref{NBARNONLINEAR}. For example, in \cite{IANNELLI1995},
solutions are attempted by controlling the analogous non-linear
integral term.  In the adder-sizer problem, we can define $\langle
\beta(t) \rangle = \int_{\Omega} \beta \bar{n}$ in the above
expression to find a self-consistent condition on $\langle
\beta(t)\rangle$. One can also assess the steady-state $\bar{n}_{\rm
  ss}$ by setting $\frac{\partial \bar{n}_{\rm ss}}{\partial t} = 0$
and establishing convergence.

One indication of blow-up is a diverging mean cell size $\langle
x(t)\rangle = M(t)/N(t)$.  By multiplying the Eq.~\ref{NBAR} by $x$
and integrating (using the boundary condition and symmetry of the
$\tilde{\beta}$ distribution) we find
\begin{equation}
{\dd \langle x(t) \rangle \over \dd t} +
\langle \beta(t)\rangle  \langle x(t)\rangle = q(t),
\end{equation}
in which $q(t)\coloneqq \int_{\Omega} g\bar{n}$.  If $\beta$, $g$, and
$\bar{n} = \bar{n}_{\rm ss}$ are time-independent and a steady state
mean cell size exists, we expect it to obey $\langle x(\infty)\rangle
= q(\infty)/\langle \beta(\infty)\rangle$.  For the special case of
deterministic exponential growth $g(x) = \lambda x$, we can write the
time evolution of the mean size as

\begin{equation}
\frac{\dd\langle x(t)\rangle}{\dd t}=
\left[\lambda - \langle \beta(t)\rangle \right] \langle x(t)\rangle, \quad 
\langle \beta(t)\rangle \equiv \int_{0}^{\infty}\!\dd x\int_{0}^{x}\!\dd y\,
\beta(x,y,t) \bar{n}(x,y,t).
\end{equation}

If $\beta(\infty)$ is bounded above by $\lambda$, then we expect
blow-up. For $\beta(\infty)$ that is not bounded, as in our example
(Eq.~\ref{BETA_GAMMA}), one cannot determine if blow-up occurs without
a more detailed and difficult analysis. Since the precise conditions
on $\beta$ leading to cell volume explosion are difficult to find, we
will explore this possible phenomena using numerical experiments. We
numerically examine the density $n(x,y,t\to \infty)$ and the mean cell
size $\langle x(t)\rangle$ using the $\beta, \tilde{\beta}$ defined in
\ref{BETA_GAMMA}, \ref{GAMMAA}, and \ref{HR}.
\begin{figure}[h!]
\begin{center}
\includegraphics[width=6.4in]{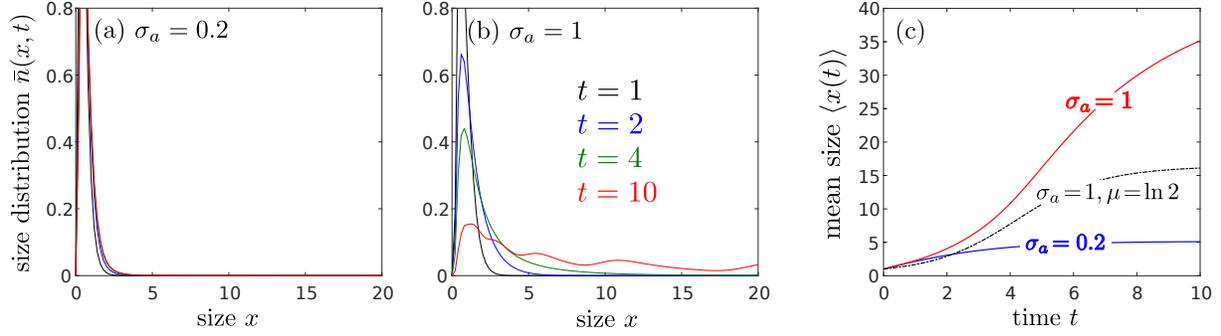}
\caption{(a) Size distributions $\bar{n}(x,t)$ for $\sigma_{a} = 0.2$
  at times $t=1,2,4,10$. (b) $\bar{n}(x,t=1,2,4,10)$ for $\sigma_{a} =
  1$, $\sigma_{r}=0.1$, and $\delta = 0$. (c) The corresponding
  mean cell sizes $\langle x(t)\rangle$.  The curve associated with
  the $\sigma_{a} = 0.2$ saturates while the one corresponding to
  $\sigma_{a}=1$ exhibits blow-up. However, the blowup is suppressed if a death 
term ($\mu=\ln2$) is included.}
\label{BLOWUP}
\end{center}
\end{figure}

In Fig.~\ref{BLOWUP}(a) and (b) we plot the marginal distribution
$\bar{n}(x,t)\coloneqq \int_{x}^{\infty}\!\dd y\, n(x,y,t)/
\int_{0}^{\infty}\!\dd x \int_{x}^{\infty}\!\dd y\, n(x,y,t)$ for
different values of the division rate variability $\sigma_{a}$ at
different times. The associated division rates correspond to those
plotted in Fig.~\ref{BETA_PLOT}(a) and (b).  In Fig.~\ref{BLOWUP}(c)
we plot the mean cell sizes $\langle x(t)\rangle = M(t)/N(t)$
corresponding to the distributions in (a) and (b).  For sufficiently
broad division probabilities $\gamma(a)$ (large $\sigma_{a}$), the
division rates $\beta$ are small, and $\langle x(t)\rangle$ fails to
saturate and diverges.

\subsection{Mother-daughter growth rate correlation}

Recent experiments indicate that the growth rate of a mother cell is
``remembered'' by its daughter cells. For growth rates of the form $g(x,y,t)
= \lambda x$, the exponential growth parameter $\lambda$ between
successive generations $i, i+1$ have been proposed to evolve
\cite{AMIR2017, DOUMIC_ESTIMATION2015}.  In \cite{AMIR2017},
fluctuations in $\lambda$ have been discussed at the single-cell level
to explore their effects on the population-averaged growth rate while
in \cite{DOUMIC_ESTIMATION2015}, changes in growth rates across two
consecutive generations are modeled as a Markov process in order to
estimate a division rate function $\beta$.  In this subsection, we
first introduce a generalized adder-sizer PDE incorporating
variability in $\lambda$ and then explore the mother-daughter growth
rate correlation affects the population dynamics.

A mother-daughter growth rate correlation between two consecutive
generations can be described by

\begin{equation}
\lambda_{i+1} = (\lambda_i-\bar{\lambda})R + \bar{\lambda} + \xi,
\label{LAMBDAD}
\end{equation}
where $\xi$ is a random variable, $0\leq R<1$ is the
successive-generation growth rate correlation, and $\bar{\lambda}$ is
the mean long-term, or preferred growth rate.  Given a growth rate
$\lambda_{i}$ of a mother cell, Eq.~\ref{LAMBDAD} describes the
predicted growth rate $\lambda_{i+1}$ of its daughter cells.  We
assume that the random variable has mean zero and is distributed
according to some probability density $P(\xi)$, which
vanishes for $\xi \leq (1-R)\bar{\lambda}$ to ensure that the growth
rates remain positive.

To incorporate the memory of growth rates between successive
generations in the adder-sizer PDE model, we extend the cell density
in the growth rate variable $\lambda$. Thus, $n(x, y, t, \lambda)$ is
the density of cells with volume $x$, added volume $y$, \textit{and}
growth rate $\lambda$. The growth function $g(x,y,t,\lambda)$ is now
explicitly a function of the growth rate $\lambda$. We propose the
extended PDE model

\begin{equation}
\left\{
\begin{aligned}
\frac{\partial n(x,y,t,\lambda)}{\partial t} +
\frac{\partial (gn)}{\partial x} &
+ \frac{\partial (gn)}{\partial y}
=-\beta(x, y,t)n(x, y, t, \lambda), \\
g(x,0,t,\lambda)n(x, 0, t, \lambda)= & 2\int_0^{\infty}\!\dd\lambda' \int_x^{\infty}\!\dd x'
\int_0^{x'}\!\dd y\, \tilde{\beta}(x', y, x,t)n(x', y, t, \lambda')
P(\xi = \lambda - R\lambda' - (1-R)\bar{\lambda}), \\
\tilde{\beta}(x, y, x', t) = &\, \tilde{\beta}(x, y, x-x',t), \\
n(x, y, 0, \lambda) = &\, n_{0}(x, y, \lambda),
\label{LAMBDA_PDE}
\end{aligned}
\right.
\end{equation}
%
%
A possible symmetric mean zero distribution that vanishes at
$-(1-R)\bar{\lambda}$ takes on a log-normal form:

\begin{equation}
P(\xi) \propto
\exp\left[-\frac{\ln^{2}(\xi+(1-R)\bar{\lambda})}{2\sigma_{\xi}^{2}}
  -\frac{\ln^{2}((1-R)\bar{\lambda} -\xi)}{2\sigma_{\xi}^{2}}\right].
\end{equation}
If we start with one newly born daughter cell at size $x_{0}$ and
growth rate $\lambda_{0}$, the initial condition in our PDE model
would be $n_{0}(x,y,\lambda) =
\delta(x-x_{0})\delta(y)\delta(\lambda-\lambda_{0})$.

\begin{figure}[h!]
\begin{center}
\includegraphics[width=6.6in]{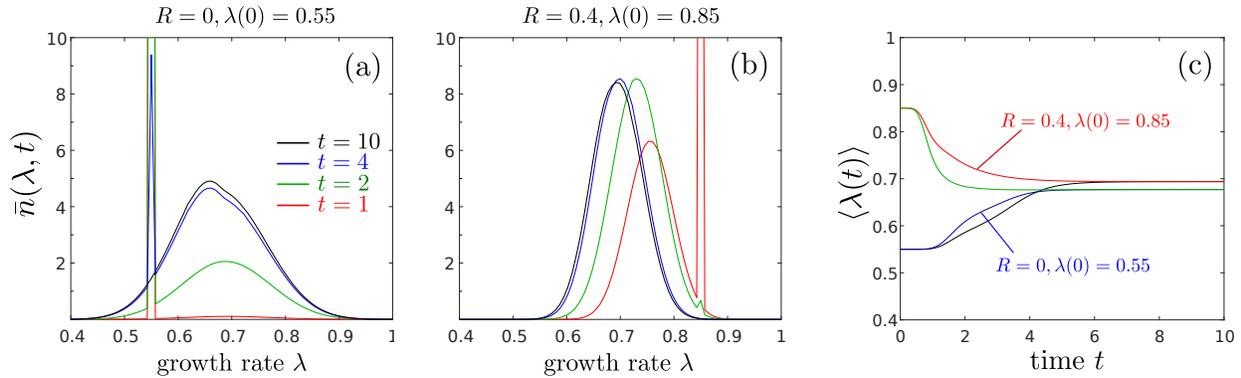}
\caption{Population-level evolution of cellular growth
  rate. Parameters used are $\bar{\lambda} = \ln 2,
    \sigma_{a}=0.2, \sigma_{r} = 0.1, \delta=0$. (a-b) The
  marginalized density $\bar{n}(\lambda, t)$ as a function of growth
  rate $\lambda$ for no correlation ($R=0$) and initial growth rate
  $\lambda = 0.55$. The peak in the distribution broadens as
    the mean evolves towards the preferred mean value
    $\bar{\lambda}=\ln 2$.  (c) The evolution of the mean $\langle
  \lambda(t)\rangle$ for different values of correlation
  $R$. Note that the steady-state values
    $\langle\lambda(\infty)\rangle$ depend on the correlation $R$.}
\label{LAMBDA}
\end{center}
\end{figure}

Numerical solutions of Eqs.~\ref{LAMBDA_PDE} shown in
Fig.~\ref{LAMBDA} indicate that although $\bar{\lambda}$ is the same
for two different cases, $R=0$ and $R=0.4$, their corresponding mean
growth rates $\langle\lambda(t)\rangle$ converge to different
values. For larger correlation $R$, the daughter cells' growth rates
do not deviate much from those of their mothers' growth rates. This
means that the offspring of faster growing cells tend to grow faster
and the offspring of slower growing cells tend to grow slower.
Because it takes shorter time for faster cells to divide, they will
produce more generations of faster-growing cells, leading to a larger
average growth rate defined as

\begin{equation}
\langle \lambda(t) \rangle = \frac{\int_0^{\infty}\dd{x}
\int_0^x\dd{y}\int_0^{\infty}\!\dd\lambda \,\, 
\lambda n(t, x, y, \lambda)}{\int_0^{\infty}\dd{x}
\int_0^x\dd{y}\int_0^{\infty}\dd\lambda n(t, x, y, \lambda)}.
\end{equation}

On the other hand, for a fixed mother growth rate $\lambda_{i}$,
smaller correlations $R$ lead to mean daughter cell growth rates
$\langle \lambda_{i+1}\rangle$ that are closer to $\bar{\lambda}$.
Since cells with growth rates less than $\bar{\lambda}$ will live
longer before division, these cells persist in the population longer
than those with larger $\lambda$, pushing the average growth rate
$\langle \lambda(t)\rangle$ to values smaller than $\bar{\lambda}$.
Fig.~\ref{LAMBDA}(c) explicitly shows that when $R=0$, the mean growth
rate approaches a value smaller than $\bar{\lambda}=\ln 2$.

\subsection{Initiation-Adder Model}

Recent experiments suggest a new type of adder mechanism for bacterial
cell size control \cite{JUN2019}.  Rather than a fixed volume added
between birth and division as the primary control parameter, new
experimental evidence suggests that the control parameter in
\textit{E. coli} is the added volume between successive initiations of
DNA replication. Initiation occurs when the \textit{ori} sites in
a cell's genome are separated, leading to DNA replication
and segregation. The number of \textit{ori} sites depend on cell type and
species, typically one in prokaryotic cells and more than one in
eukaryotic cells.  The initiation-adder model assumes that a cell's
volume per initiation site (the \textit{ori} site in the genome) tends
to add a fixed volume between two consecutive initiations.

If the number of \textit{ori} sites in a cell is $q$, initiation
increases the number to $2q$.  Immediately after division and DNA
separation, the number of \textit{ori}s decreases back to $q$ in each
daughter cell.

\begin{figure}[h!]
\begin{center}
\includegraphics[width=3.75in]{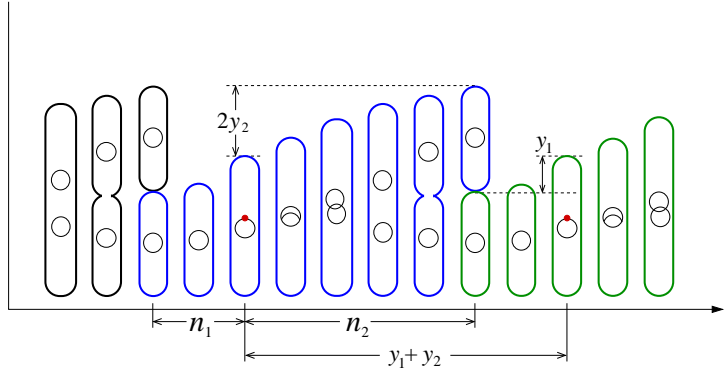}
\caption{Schematic for the initiation adder process.  DNA
    replication is initiated (indicated by the red dot) before copied
    DNA is segregated and cell division. In this example, $q=1$ and
    $y_{2}$ is and added volume per origination site for two
    origination sites. The density of cells with $q=1$ copy of DNA
    (before DNA replication initiation) is denoted $n_{1}(x,y,t)$
    while the density of cells post-initiation is denoted
    $n_{2}(x,y,t)$, where $y$ denotes the volume added after
    initiation. The factor that controls $y_{1}+y_{2}$ in the
    initiation-adder model is the volume $\Delta$ added between
    successive initiation events, rather than between successive cell
    divisions. Thus, the controlled variable (added volume in this
    case) spans the pre-initiation and post-initiation states.}
\label{INITIATION}
\end{center}
\end{figure}

In this subsection, we generalize the adder PDE model to describe this
new initiation-adder mechanism. We classify all cells into two
subpopulations: cells that have not yet undergone initiation and cells
that have initiated DNA replication but that have not yet divided.  We
define $n_1(x, y, t)\dd{x}\dd{y}$ as the expected number of
pre-initiation cells in with volume in $[x,x+\dd{x}]$ and with added
volume $y<x$ in $[y, y+\dd{y}]$.  Mean post-initiation cell numbers
with volume in $[x,x+\dd{x}]$ and added volume in $[y, y+\dd{y}]$ are
described by $n_{2}(x, y, t)\dd{x}\dd{y}$. In the general
initiation-adder process, when a pre-initiation cell commences DNA
replication (initiates) can depend on the volume or added
volume. Thus, we describe transitions from a pre-initiation cell
transitions into a post-initiation cell by the rate $k_{\rm
  i}(x,y,t)$.  After initiation, the number of \textit{ori} sites
doubles and the added volume is reset to zero in the newly formed
post-initiation cell. In analogy with the differential division rate
in Eq.~\ref{AS_PDE}, we define $\beta(x,y,t)$ as the rate of division of
post-initiation cells.  Under a general asymmetric division event, we
assume that the added volume is divided proportionally to the volume
of the daughter cells, \textit{i.e.}, if the mother cell's volume is
$x$ with added volume $y$ since initiation, and if one daughter cell's
volume is $z<x$ and the other daughter cell's volume is $x-z$, the
added volume since division for the first daughter will be set to
$yz/x$ while the added volume for the second daughter will be
$y(x-z)/x$.  The resulting PDE model now involves two coupled
densities $n_{1}$ and $n_{2}$:

\begin{equation}
\begin{aligned}
\frac{\partial n_1(x,y,t)}{\partial t} 
+ \frac{\partial [g_{1} n_1]}{\partial x} + 
\frac{\partial [g_{1}n_1]}{\partial y} &  
= -k_{\rm i}(x,y,t)n_{1} + 2\int_x^{\infty}\frac{z}{x}n_2(z,yz/x,t)
\tilde{\beta}(z, x, yz/x,t)\dd{z}, \\
\frac{\partial n_2(x, y, t)}{\partial t} + 
\frac{\partial [g_{2}n_{2}]}{\partial x}+ \frac{\partial [g_{2}n_{2}]}{\partial y} &  
= -\beta(x,y,t)n_{2}, \\
n_{1}(x,0,t) = 0, \quad g_{2}n_2(x, 0, t) & 
= \int_0^x k_{\rm i}(x,y,t)n_1(x, y, t)\dd{y}, \\
\beta(x, y, t) & = \int_0^x\tilde{\beta}(x, z, y, t)\dd z,
\label{IAEQNS}
\end{aligned}
\end{equation}
in which we have allowed for different growth rates in the different
cell phases. Both $n_1$ and $n_2$ are defined in the domain
$\{{\mathbb{R}^+}^2\cap\{y<x\}\}\times\mathbb{R}^+$.  These coupled
PDEs are different from the PDE associated with the standard
``division-adder'' described in Eqs.~\ref{AS_PDE} and \ref{AS_BC}.
Here, the added volume is reset to zero not after division, but after
initiation.

In \cite{Wallden2016}, a strong size control acting on initiation
initiation was proposed where all cells will have inititated DNA
replication before reaching some fixed volume $x_{\rm i}$. This
hypothesis can be implemented in our initiation-adder model by setting
$k_{\rm i}(x\to x_{\rm i}, y, t)\to \infty$. The probability that a
cell born at time $t_0$ has not yet initiated, $e^{-\int_{t_0}^t
  k_{\rm i}(x(s), y(s), s)\dd{s}}$, always vanishes for all $(t_0,
x_{t_0}, y_{t_0})$ before some finite time $t$ and $x(t)< x_{\rm i}$.
Thus, $n_2(x, 0, t)$ is nonzero only in $[0, x_{\rm i}]$ for all $t$.
If there exists a constant $\tau_0$ such that
$\lim\limits_{\tau\rightarrow\tau_0} e^{-\int_{t_0}^{t_0+\tau}\!k_{\rm
    i}(x(s), y(s), s)\dd{s}}=0$ for all $t_0$, then the largest volume
that any cell can attain will be $e^{\lambda{\tau_0}}x_{\rm i}$,
leading to strict size control and no blowup.
\begin{figure}[h!]
\begin{center}
\includegraphics[height=3.7in]{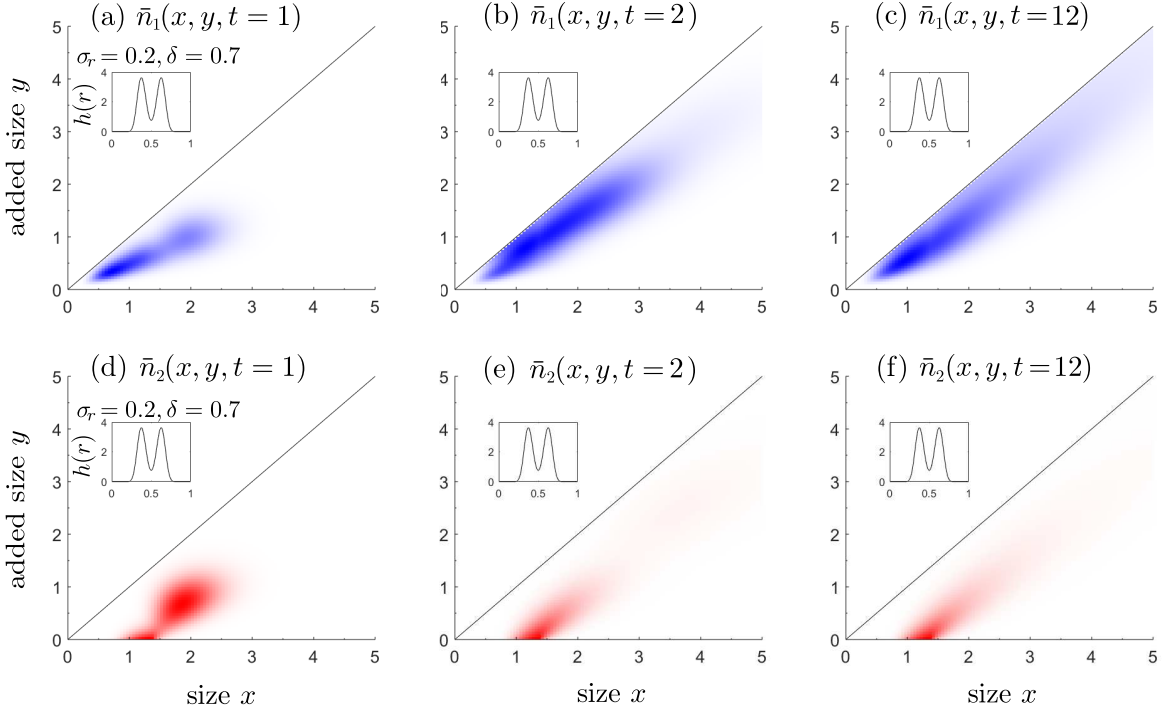}
\caption{Normalized densities of pre-initiation cell
    populations $\bar{n}_{1}$ and post-initiation cell populations
    $\bar{n}_{2}$ for at various fixed times $t=1,2,12$. Here, we used
    $k_{\rm i}(x) = p(x)/\left[1-\int_{0}^{x}p(x')\dd x'\right]$ with
    $p(x) \sim {\cal N}(1,0.1)$ and the same $\tilde{\beta}(x,y,z,t)$
    as that used in Fig.~\ref{DENSITY}(d-f).  (a-c) shows the
    normalized densities $\bar{n}_{1}(x,y,t)\equiv n_{1}(x,y,t)/N(t)$
    where $N(t) = \int\dd y \int\dd x (n_{1}+n_{2})$. (d-f) shows the
    normalized post-initiation density $\bar{n}_{2}(x,y,t)$. For the
    $k_{\rm i}$ used in this example, the pre-initiation densities
    span larger volume and added volumes. The densities are
    indistinguishable from those at steady state after about $t =
    2$.}
\label{IA}
\end{center}
\end{figure}

Fig.~\ref{IA} shows numerical solutions to Eq.~\ref{IAEQNS} using the
same birth rate function as that used in Fig.~\ref{DENSITY}(d-f). Note
that due to cell size control affecting the pre-initiation stage,
initial daughter cell sizes stay small at initiation and
$n_{1}(x,y,t)$ is more peaked near $y \approx x$.

If one takes $k_{\rm i}$ sufficiently large, both daughter cells will
nearly instantly initiate DNA replication after division. We have
checked numerically that for constant $k_{\rm i}=10^3$, the densities
$n_{1}(x,y,t)$ are negligible while $n_{2}(x,y,t)$ approaches the
density of the division adder shown in Fig.~\ref{DENSITY} (for the
same differential division functions $\tilde{\beta}$). Thus, the
initiation adder model converges to the standard division adder model
when $k_{\rm i}\to \infty$.  This can be seen from the first of
Eqs.~\ref{IAEQNS} where $n_{1}$ can be neglected and is dominated by
the two terms on the right-hand-side. Substituting $k_{\rm
  i}(x,y,t)n_{1} \approx 2\int_x^{\infty}\frac{\dd z}{x}n_2(z,yz/x,t)$
into the integral terms in the second equation, we find
Eq.~\ref{AS_PDE} for $n_{2}(x,y,t)$.

\section{Summary and Conclusions}

In this paper, we used PDE models to describe population dynamical
behavior under the adder division mechanism. Under certain conditions,
this PDE for the adder mechanism can also be converted to the
well-known size- and age-structured PDE.  In the absence of death, we
motivated models for the differential birth rate function
$\tilde{\beta}(x,y,z,t)$ that are consistent with normalized division
probabilities In Appendix A we showed existence and uniqueness of a
weak solution to the PDE model within a time interval $[0, T]$ during
which the solution's support can be bounded. One can prove similar
results when both time and space are unbounded as this problem is
related to other first order PDE models that have been studied in more
detail.

With a weak solution justified, we explored the ``adder-sizer'' PDE
via numerical experiments and Monte-Carlo simulations of the
underlying stochastic process. Our results show that event-based
Monte-Carlo simulations of the discrete process generate sample
configurations. The observed configurations are are consistent with
samples from the cell densities numerically computed from our PDE
model.

When broader differential division rates are used (when cell division
is more asymmetric), we find, under the same initial conditions, a
broader cell density $n(x,y,t)$ and a broader event density $\rho_{\rm
  D}(x,y,T)$. We also demonstrate numerically, the divergence of the
mean cell size $\langle x(t)\rangle = M(t)/N(t)$.  We showed
that division probabilities that are broader in the age or added size
(and smaller in magnitude) more likely lead to mean cell sizes that
explode with time. 

We then incorporated growth rate correlation between cells of
successive generations \cite{KESSLER2017} into our ``adder-sizer'' PDE
model. By extending the dimension of the density function to include
growth rates and allowing for variability in growth rate as new cells
are born, we developed a PDE model that incorporated the stochastic
nature of growth rate inheritance and that describes evolution of the
growth rate distribution of cells. We found that the steady-state
value of the mean growth rate depends on the correlation of growth
rates between mother and daughter cells. This dependence arises from a
subtle interaction between the shape of the growth rate distribution
and the distribution of variations in the growth rate from one
generation to the next.

Finally, we proposed a coupled partial integro-differential equation
(PIDE) to model two-phase cell population dynamics under a new
initiation-adder mechanism suggested by recent experimental
results. In the limit that the initiation rate $k_{\rm i}$ of DNA
replication is significantly faster than all other time scales in the
problem, the numerical solutions of the initiation adder model
(Eq.~\ref{IAEQNS}) converge to those of division adder model
(Eqs.~\ref{AS_PDE} and \ref{AS_BC}). Under proper assumptions that
come from experimental findings, we found that the initiation adder
would also lead to effective cell size control \cite{Wallden2016}.

There are new cellular processes and size control mechanisms that have
been recently discovered and that can be mathematically modeled.
Thus, there is likely general mathematical topics that remain to be
explored within PDE and PIDE models of structured populations.  For
example, a recent experimental study indicates that an adder mechanism
may be the result of several consecutive processes in the cell
division cycle, suggesting that a much more complicated coupled system
of PDEs/PIDEs would be required.

\section*{Acknowledgments}
This work was supported in part by grants from the NSF (DMS-1814364),
the National Institutes of Health (R01HL146552) and the Army Research
Office (W911NF-18-1-0345).  
 


\appendix
\section{Existence and uniqueness of a weak solution for the adder-sizer model}
\label{A}

In this section we show the existence and uniqueness of the solution to the
``adder-sizer'' model PDE. The full problem is defined as 

\begin{equation}
\left\{
\begin{aligned}
\frac{\partial{n}}{\partial{t}} + \frac{\partial{(ng)}}{\partial{x}}
+ \frac{\partial{(ng)}}{\partial{y}} =-\beta(x, y, t) n(x, y, t),\\
g(x,0,t)n(x, 0, t)=2\int_x^{\infty}\!\!\dd x'\int_0^{x'}\!\!
\dd{y}\, \tilde{\beta}(x', y, x, t)n(x', y, t),\\
\beta(x, y, t) \coloneqq \int_0^{x}\tilde{\beta}(x, y, z, t)\dd z,\\
\tilde{\beta}(x, y, z', t) = \tilde{\beta}(x, y, z-z', t),\,\, 
\tilde{\beta}(x, y, 0, t) = 0,\,\, n(x, x, t) =0,\\
n(x, y, t=0) \coloneqq n_0(x, y).
\end{aligned}
\right.
\label{PDEForm}
\end{equation}
where the independent variables  $(x, y,t)\in
\mathbb{R}^2\cap\{y<{x}\}\times \mathbb{R}^+$.

First, we assume that

\begin{equation}
\begin{aligned}
0 < g_{\min}\leq{g}\in & \textbf{C}^{1}(\{(\mathbb{R}^+)^2\cap\{y\leq{x}\}\}\times\mathbb{R}^+), \\
n_0(x, y) \in  & \textbf{L}^1\cap\textbf{L}^{\infty}\cap
\textbf{C}^{1}({\mathbb{R}^+}\cap{\{y<x\}}), \\
0 \leq \tilde{\beta} \in & \textbf{L}^{\infty}\cap\textbf{L}^1\cap
\textbf{C}^{1}(\{(\mathbb{R}^+)^3\cap{\{y<x, z<x\}}\}\times\mathbb{R}^+), \\
\beta(x, y, t) \in & \textbf{L}^{\infty}
\cap\textbf{L}^{1}\cap
\textbf{C}^{1}(\{(\mathbb{R}^+)^2\cap\{y\leq{x}\}\}\times\mathbb{R}^+),
\label{A2}
\end{aligned}
\end{equation}
and nondimensionalize the size and added size by $\Delta$,
the added size parameter defined in Eq.~\ref{A_XY}.
We also impose an additional assumption on $g$:

\begin{equation}
|g(x, y, t)| < K(t + x + 1),\quad  K < \infty.
\label{gBound}
\end{equation}
We also assume the initial distribution $n_0(x,y)$ is compactly
supported in $(0, \Omega)\times[0, \Omega), \Omega<\infty$.  From this
  assumption and \ref{gBound}, the closure of $n(x, y, T)$'s support
  is compact for any finite time $T$ since $n\neq 0$ only when $y<x$
  and $x(s) \leq Ce^{Ks} - (1+T)$
%
from Gr\"{o}nwall's Inequality, where $C<1+T+\Omega$ is given by the
initial condition. At any finite time $T$, the support of $n(x, y,T)$
is bounded and we assume it is contained in $[0, \Omega(T))\times[0,
    \Omega(T))$. Furthermore, by setting $g, \beta, \tilde{\beta}=0$
    at the given time $T$ when $(x, y)$ is out of the support of $n$,
    we can assume the closure of $g, \beta, \tilde{\beta}$'s support
    to be compact. One can generalize the definition of the weak
    solution $n$ to $[(\mathbb{R}^+)^2\cap\{y<x\}]\times [0,
      \infty)$ as in \cite{PERTHAME2008}.

\textbf{Definition A.1} Given time $T<\infty$ and assuming \ref{A2}, for
a function $n\in\textbf{L}^1((([0,
  \Omega(T)])^2\cap\{y<x\})\times[0, T]), \Omega(T)<\infty$ with $n(x,
y, t)\neq0$ in $[0, \Omega(T))\times[0, \Omega(T)), y<x, t\in[0, T]$, $n$ is said to 
    satisfy the adder-sizer PDE in the weak sense in time $[0, T]$,
    if 
\begin{equation}
\begin{aligned}
-\int_0^T\!\!\dd t\int_0^{\infty}\!\!\dd x\int_0^{x}\!\!\dd y\, n(x,
y, t)\left[\frac{\partial{\Psi}}{\partial{t}}+g(x,y,t)
\frac{\partial{\Psi}}{\partial{x}}+g(x,y,t)\frac{\partial{\Psi}}{\partial{y}}
-\beta(x,y,t)\Psi(x,y,t)\right] \\
=\int_0^{\infty}\!\!\dd x\int_0^{x}\!\!\dd y\, n_0(x, y)\Psi_0(x, y) +
\int_0^T\!\!\dd t\int_0^{\infty}\!\!\dd x\, \Psi(x, 0, t)n(x, 0, t)g(x, 0, t)
\label{AdjPro}
\end{aligned}
\end{equation}
holds for all test functions $\Psi\in\textbf{C}^1(([0,
      \Omega(T)])^2\cap\{y\leq{x}\})\times[0, T])$ satisfying $\Psi(x,
    y, T)\equiv0, \Psi(\Omega(T), y, t)=0$ and $\Psi(x, x, t)=0$, 
where we set $g, \tilde{\beta}, \beta=0$ for $x \geq \Omega(T),
x\leq{y}$ or $x\leq{z}$. Upon using the boundary condition in
\ref{PDEForm}, the right-hand-side becomes

\begin{eqnarray*}
\begin{aligned}
\int_0^{\infty}\!\!\dd x\int_0^{x}\!\!\dd y\, n_0(x, y)\Psi_0(x, y) +
2\int_0^T\!\!\dd t \int_0^{\infty}\!\!\dd x\int_{0}^{x}\!\!\dd y\int_{0}^{x}\!\!\dd z\,
\Psi(z,0,t)\tilde{\beta}(x,y,z,t)n(x,y,t).
\end{aligned}
\end{eqnarray*}

Note that if
$n\in\textbf{C}^1(((\mathbb{R}^+)^2\cap\{y<x\})\times\mathbb{R}^{+})$
is a classical solution to the PDE (Eq.~\ref{PDEForm}), then it must
also satisfy Eq.~\ref{AdjPro} in any time interval $[0, T]$.  We refer
to \cite{PERTHAME2008} for a proof of the existence and uniqueness of
a weak solution of a related, simpler renewal equation.  However, our
adder-sizer PDE is more complicated. The proof of uniqueness requires
very different techniques from the sizer PDE; yet the proof of
existence is similar to the proof in \cite{PERTHAME2008}.

\subsection{Uniqueness}
First, we prove uniqueness of the solution to \ref{AdjPro}. Assume
there are two weak solutions $n^{(0)}$ and $n^{(1)}$ for the adder-sizer PDE
satisfying \ref{AdjPro} with the same initial condition
$n_0^{(0)}(x, y)=n_0^{(1)}(x, y)$. Taking the difference between using these
purported solutions, we obtain 

\begin{equation}
\begin{aligned}
-\int_0^T\!\!\dd t\int_0^{\infty}\!\!\dd x\int_0^{x}\!\!\dd y\, \Delta n(x,y,t)
\left[\frac{\partial{\Psi}}{\partial{t}}+g(x,y,t)
\frac{\partial{\Psi}}{\partial{x}}+g(x,y,t)
\frac{\partial{\Psi}}{\partial{y}}-\beta(x,y,t)\Psi(x,y,t)\right]\\
=2\int_0^T\!\!\dd t \int_0^{\infty}\!\!\dd x \int_{0}^{x}\!\!\dd y \int_{0}^{x}\!\!\dd z\,
\Psi(z, 0, t)\tilde{\beta}(x,y,z,t)\Delta n(x,y,t),
\label{DeltePsi}
\end{aligned}
\end{equation}
where $\Delta n = n^{(1)} - n^{(0)}$.

\subsubsection{Adjoint Problem}
We consider the adjoint problem for $\Psi$ in the given time interval
$[0, T]$ and with a source term $S(x,y,t)$:

{\small\begin{equation}
\begin{aligned}
\frac{\partial{\Psi}}{\partial{t}}+g(x,y,t)
\frac{\partial{\Psi}}{\partial{x}}+g(x,y,t)
\frac{\partial{\Psi}}{\partial{x}}-\beta(x,y,t)\Psi(x,y,t) 
&= -2\int_0^x\Psi(z, 0, t)\tilde{\beta}(x, y, z, t) \dd{z} - S(x,y,t), \quad
0\leq{y} < x \\ 
\Psi(x,y,T)& =0, \,\, \Psi(\Omega(T), y, t)=0, \,\, \Psi(x,x,t)=0.
\label{AdjPro2}
\end{aligned}
\end{equation}}

\textbf{Theorem A.1} Assume \ref{A2}, and $S\in\textbf{C}^1([0, \Omega(T)]^2\times[0,
  T]), S(\Omega(T), y, t)=0$, and $S=0$ when
$x\leq{y}$. Then there exists a unique $\textbf{C}^1$ solution to the
adjoint problem.

\textbf{Proof.} We can transform the above equation into an ODE along
the characteristic line and use contraction mapping, which is a
standard practice in functional analysis to prove existence and
uniqueness of the solution to a PDE problem. On the left-hand-side of
Eq.~\ref{AdjPro2}, we apply the characteristic line method. Setting
$X(c,t)=(x(c, t),y(c,t))$ on the characteristic lines leads to

\begin{equation*}
\left\{
\begin{aligned}
\frac{\partial{X(c, s)}}{\partial{s}}=(g(x, y, s), g(x, y, s)), \quad 
t\leq{s}\leq{T},\\ X(c, t)=(x_t, y_t), \quad 0\leq{y_t}<x_t, x_t-y_t=c.
\end{aligned}
\right.
\end{equation*}
Since we have $x(s)-y(s)=x_t-y_t$, the above equation can be
simplified to

\begin{equation*}
\frac{\partial{X}(c, s)}{\partial{s}}=
\tilde{g}(X(c, s), s),\,\, x(c, t)=x_t,\,\, y(c, t) = x_t - c
\end{equation*}
where $\tilde{g}(X(c, s), s)=(g(x(c, s), x(c, s)-c, s),\, g(x(c, s),
x(c, s)-c, s))$. Once $c$ is fixed and $x_t$ is given, the above
equation becomes an ordinary differential equation. Given $x_t$, we
define

\begin{equation*}
\left\{
\begin{aligned}
\tilde{\Psi}(c, s)\coloneqq \Psi(X(c, s), s)e^{-\int_{t}^{s}\beta(X(c, v), v) dv},\quad\quad\quad\quad\quad\quad\quad\quad\quad\quad\quad\quad\quad\quad\quad\quad\quad\quad\quad\quad\\ 
U(c, z, s)\coloneqq 2\tilde{\beta}(X(c, s), z, s)
e^{-\int_t^s\beta(X(c, v), v) dv}, \,\, \tilde{S}(c, s)\coloneqq S(X(c, s), s)e^{-\int_t^s\beta(X(c, v), v) dv}.
\end{aligned}
\right.
\end{equation*}
Thus, along the characteristic line we can write \ref{AdjPro2} as
\begin{equation}
\frac{\partial}{\partial{s}}\tilde{\Psi}(c,s)
=-\int_{0}^{x(c,s)}\Psi(z, 0, s)U(c, z, s)\dd z-\tilde{S}(c,s).
\end{equation}
Since $\tilde{\Psi}(c, T)=0$ and $\tilde{\Psi}(c, t)=\Psi(x_t,x_t-c, t)$,

\begin{equation}
\Psi(x_t, x_t-c, t)=\int_t^{T}\tilde{S}(c, s)\dd s +\int_t^{T}\!\!\dd s
\int_0^{x(c, s)}\!\!\dd z\, \Psi(z, 0, s)U(c, z, s),\quad  0 < c\leq{x_t}.
\label{PSI0}
\end{equation}

We can see that if $x\leq y$ or $x_t \geq \Omega(T)$, $\Psi(t, x_t,
x_t-c)= \Psi(t, x, x)=0$ since $U, \tilde{S}=0$ for $c\leq0$ or
$x_t>\Omega(T)$. Using $c=x_t$, Eq.~\ref{PSI0} becomes

\begin{equation}
\Psi(x_t, 0, t)=\int_t^{T}\! \tilde{S}(x_t, s)\dd s + \int_t^T\!\!\dd s
\int_0^{x(x_t, s)}\!\!\!\!\dd z\, \Psi(z, 0, s)U(x_t,z,s).
\end{equation}
From condition \ref{gBound} we obtain
$x(s)\leq(x_t+1+T)e^{K(s-t)}-(1+T)$. From condition A.3, we define
$\tilde{B} = 2\|\tilde{\beta}\|_{\infty} < \infty$.
Next, we choose $s=\max\{T-\frac{1}{K}\ln(1+\frac{1}{2\tilde{B}(1+T)}),
T-\frac{1}{K}\ln2, T-1\}$ such that
$e^{K(T-t)}\leq1+\frac{1}{2\tilde{B}(1+T)}, s\leq{t}\leq{T}$, and choose $x_s$
small enough such that $x_s<\min\{1, \frac{1}{8\tilde{B}(T-s)}\}$. We
denote a mapping $T$ defined on the functional space as

\begin{equation*}
T(\Psi)(x_t, 0, t) = \int_t^{T}\tilde{S}(x_t, s)\dd{s}+\int_t^{T}\!\!\dd s \int_0^{x(s,
  x_t)}\!\!\!\!\dd z\, \Psi(z, 0, s)U(x_t,z,s), \quad t\in[s, T], x_t\in[0,
  x_s].
\end{equation*}
It is easy to verify that $T$ is a contraction mapping for $\Psi(x_t,
0, t)$ and thus there exists a unique solution $\Psi_0$ satisfying
\ref{AdjPro2} in $D_0$ defined as $D_0=\{(x, t)|s\leq{t}\leq{T},
0\leq{x}\leq{x( x_s, t)}\}$. We then let $x_s^1>x_s$ and define $D_1=
\{(x, t)|s\leq{t}\leq{T}, 0\leq{x}\leq{x(x_s^1, t)}\}$ such that the
difference of the area between regions $D_1$ and $D_0$ is less than
$\tilde{B}^{-1}$. Next, define a second mapping $T_1$ by

\begin{equation*}
\left\{
\begin{aligned}
T_1(\Psi)(x_t, 0, t) = \int_t^{T}\!\!\dd s \int_{x(x_s, s)}^{x(x_t, s)}\!\!\!\!\dd z\, \Psi(
z, 0, s) U(x_t, z, s) + I(x_s, t), \quad  t\in[s, T],\,\,  x_t\in[x(t,
  x_s), x_s^1],\\
I(x_s, t)=\int_t^{T}\!\!\dd s\,\tilde{S}(x_t, s) +\int_t^T\!\!\!\dd s 
\int_0^{x(x_s, s)}\!\!\!\!\!\!\!\dd z\,\,
\Psi_0(z,0,s) U(x_t, z, s).\qquad\qquad\qquad\qquad\qquad\qquad\quad 
\end{aligned}
\right.
\end{equation*}
$T_1$ is also a contraction mapping and we can obtain a $\Psi_1$ on
$D_1$ such that $T(\Psi_1)=\Psi_1$. Denote

\begin{equation}
\Psi(x, 0, t)=\left\{
\begin{aligned}
\Psi_0(x, 0, t),\; (x, t)\in{D}_0,\\
\Psi_1(x, 0, t),\; (x, t)\in{D}_1,
\end{aligned}
\right.
\end{equation}
and it is easy to verify that $\Psi$ is $C^1$ continuous on
$D_0\cap{D_1}$ by first proving it is continuous and then taking the
partial derivatives, and $\Psi$ satisfy \ref{AdjPro2} in the region
$D_0\cup{D_1}$.

Following the same procedure, we can extend $\Psi$ to satisfy
\ref{AdjPro2} in the region $t\in[s, T]$. Then, for $[0, s]$, we
choose a $\tilde{s}$ close enough to $s$ and use the same strategy by
defining $T_2$ as

\begin{equation}
\left\{
\begin{aligned}
T_2(\Psi)(x_t, 0, t)= \int_t^{s}\!\!\dd r\, \tilde{S}(x_t, r)+
\int_t^s\!\!\dd r \int_0^{x(x_t, r)}\!\!\!\!\dd z\, \Psi(z, 0, r) U(x_t, z, r)
+\tilde{I}(t, x_s), \,\, t\in[\tilde{s}, s], \\
\tilde{I}(x_s, t) = \int_s^T\!\!\dd r\, \tilde{S}(x_t, r)
+\int_s^T\!\!\dd r \int_0^{x(x_t, r)}\!\!\!\!\!\!\!\! \dd z\,
\Psi(z, 0, r) U(x_t, z, r)\qquad\qquad\qquad\qquad\qquad\quad.
\end{aligned}
\right.
\end{equation}
We finally obtain a unique function $\Psi$ satisfying \ref{AdjPro2} in $[0,
  T]\times[0, \infty)$.

From \ref{PSI0}, the value of $\Psi$ is determined
by $\tilde{S}, \Psi(x, 0, t), U$ and we conclude that there exists a 
unique $\textbf{C}^1$ solution for \ref{AdjPro2}.

\subsubsection{Uniqueness of weak solution for the adder-sizer model}
From Section A.1.1 we obtain the existence and uniqueness of $\Psi$
of the adjoint problem. Given any time $T$ and $S(x,
y, t)\in\textbf{C}^1(\mathbb{R}^+\times(\mathbb{R}^+)^2)$ satisfying the
condition in Theorem A.1, since we can set $g,
\beta, \tilde{\beta}$'s support to be compact in $[0, T]$, we can
find a unique $\textbf{C}^1$ continuous $\Psi$ satisfying \ref{AdjPro2}.
By substituting \ref{AdjPro2} into \ref{DeltePsi}, we 
obtain

\begin{equation}
\int_0^T\!\!\dd t \int_0^{\Omega(T)}\!\!\!\!\!\! \dd x \int_0^x\!\!\dd y\,
\Delta n(x,y,t)S(x,y,t) = 0
\end{equation}
for any $S(x,y,t)\in\textbf{C}^1(\mathbb{R}^+\times{\mathbb{R}^+}^2)$
satisfying $S(x \leq y,t)=S(x\geq\Omega(T), y, t)=0$, which implies
$n\equiv0$ a.e. in $y<x\leq{\Omega(T)}$. So at any given time $T$ the
weak solution, if exists, is unique.

One can also set the condition for $\tilde{\beta}, g$ weaker even
when we define the weak solution in unbounded region $[0,
\infty)\times(\mathbb{R}^+)^2\cap\{y<x\}$. In \cite{PERTHAME2008} such
work is done for the renewal equation. We do not discuss this
generalization in detail here.

\subsection{Existence of the weak solution}
We construct a series of functions $\{n_i\}$ with a limit $n$ for this
series satisfying \ref{AdjPro2} for all test functions $\Psi$. We use
a semi-discrete approximation to discretize the PDE and obtain
piecewise solutions. As the mesh size becomes smaller, we expect the
piecewise solution to converge to a function $n$ satisfying
\ref{AdjPro}.  The idea of constructing a series of piecewise constant
solutions and proving their convergence to a weak solution is similar
to that in \cite{PERTHAME2008}.

\subsubsection{Semi-discrete approximation for the PDE}
We choose a uniform grid with mesh size $h>0$
fixed in both $x$ and $y$ axis and let time $t$ be continuous. We
denote

\begin{equation}
\begin{aligned}
(x_i, y_j)=(ih, jh), (x_{i+\frac{1}{2}},  y_{j+\frac{1}{2}})=
((i+\frac{1}{2})h, (j+\frac{1}{2})h), \quad  j< i \in\textbf{N}, \\
\beta_{i+\frac{1}{2}, j+\frac{1}{2}}(t) = \frac{1}{h^2}\int_{ih}^{(i+1)h}\!\!\!\dd y 
\int_{jh}^{(j+1)h}\!\!\!\dd x\,  \beta(x, y, t), \quad j<i \in\textbf{N},\\
\tilde{\beta}_{i+\frac{1}{2}, j+\frac{1}{2}}((s+\frac{1}{2})h, t) =
\frac{1}{h^3}\int_{ih}^{(i+1)h}\!\!\!\!\dd z \int_{jh}^{(j+1)h}\!\!\!\!\dd y 
\int_{sh}^{(s+1)h}\!\!\!\!\dd x\, 
\tilde{\beta}(x,y,z,t), \quad s\leq{i},\\
g_{i, j}(t)=g(ih, jh, t), \quad j<i \in\frac{1}{2}\textbf{N}.
\end{aligned}
\end{equation}

Here, $\beta_{i+\frac{1}{2},
  j+\frac{1}{2}}(t)=h\sum_{s=0}^i\tilde{\beta}_{i+\frac{1}{2},
  j+\frac{1}{2}}((s+\frac{1}{2})h,t)$. Given a fixed time $T$, we wish
to find a solution of pointwise function $n^h(t)$, which takes values
on the grid points $(x_{i+\frac{1}{2}}, y_{j+\frac{1}{2}})$. Then
$n^h$ can be seen as a vector function. According to our assumption
there exists $\Omega$ such that the initial value $n^0$ is nonzero
within the region $\{(x, y)|y<x, x<\Omega\}$, and from our previous
calculation there exists $\Omega(T)<\infty$ such that $n$ is nonzero
within the region $\{(x, y)|y<x, x<\Omega(T)\}$.  Eventually, we will
set $h(k)=\Omega(T)/k$ and let the mesh size $h\to 0$ by letting $k
\to \infty$.

By discretizing Eqs.~\ref{PDEForm}, we expect the vector function
$n^h(t)$ to satisfy the below equations for $t\in[0, T]$ and $0<j<i<L$
($L$ is the number of discretization grid points along one direction):

\begin{equation}
\begin{aligned}
\displaystyle \: & h^2\frac{\dd n_{i+\frac{1}{2}, j+\frac{1}{2}}(t)}{\dd t}+h(g_{i+1, j+\frac{1}{2}}(t)
n_{i+\frac{1}{2}, j+\frac{1}{2}}(t)-g_{i, j+\frac{1}{2}}(t)n_{i-\frac{1}{2}, j+\frac{1}{2}}(t))  \\
\: & \hspace{5mm}\displaystyle + h(g_{i+\frac{1}{2}, j+1}(t)n_{i+\frac{1}{2}, j+\frac{1}{2}}(t)
-g_{i+\frac{1}{2}, j}(t)n_{i+\frac{1}{2}, j-\frac{1}{2}}(t)) + h^2\beta_{i+\frac{1}{2}, j+\frac{1}{2}}(t)
n_{i+\frac{1}{2}, j+\frac{1}{2}}(t)=0, \quad 0\leq{j}<i-1 \\[12pt]
\: & \displaystyle h^2\frac{\dd n_{i+\frac{1}{2}, j+\frac{1}{2}}(t)}{\dd t} 
+ hg_{i+1, j+\frac{1}{2}}(t)n_{i+1, j+\frac{1}{2}}(t)  \\
\: & \hspace{1cm} -hg_{i+\frac{1}{2}, j}(t)n_{i+\frac{1}{2}, j-\frac{1}{2}}(t) 
+ h^2\beta_{i+\frac{1}{2}, j+\frac{1}{2}}(t)
n_{i+\frac{1}{2}, j+\frac{1}{2}}(t)=0, \quad 0\leq{j}=i-1 \\[12pt]
\: & \displaystyle g_{i+\frac{1}{2}, 0}(t) n_{i+\frac{1}{2},
  -\frac{1}{2}}(t) = 2h^2\sum_{\ell=i}^{L-1}
\sum_{j=0}^{\ell-1}\tilde{\beta}_{\ell+\frac{1}{2}, j+\frac{1}{2}}
((i+\frac{1}{2})h,t) n_{\ell+\frac{1}{2},
  j+\frac{1}{2}}(t),  \\[12pt]
\: & \displaystyle n_{i+\frac{1}{2},
  j+\frac{1}{2}}(0)=\frac{1}{h^2}\int_{x_i}^{x_{i+1}}\!\!\!\dd y
\int_{y_j}^{y_{j+1}}\!\!\!\dd x\, n_{0}(x, y), \quad n_{i+\frac{1}{2}, i+\frac{1}{2}}(t)=0,
\label{DisForm}
\end{aligned}
\end{equation}

\noindent where we henceforth omit the $h$ superscript in the proof.
In the two-dimensional upwind scheme, derivatives in one direction are
neglected on neighboring sites in the other direction:
$n_{i,j\pm\frac{1}{2}}=n_{i-\frac{1}{2}, j\pm\frac{1}{2}},
n_{i\pm\frac{1}{2}, j}=n_{i\pm\frac{1}{2}, j-\frac{1}{2}}$.  The
boundary condition $n(x, x, t)=0$ is implemented by $n_{i+\frac{1}{2},
  i+\frac{1}{2}}(t)=0$ for any $t$ and $i$.

We will obtain a uniform bound irrelevant of $h$ for $n$.  All
coefficients in the above ODE equations are $\textbf{C}^1$ continuous,
which means that there exists a unique solution in time $[0, T],
T<\infty$.

\textbf{Theorem A.2} For $t\in[0, T]$ and assuming \ref{A2} holds, 
we find the bound
\begin{equation}
\sum_{i=1}^{L-1}\sum_{j=0}^i|n_{i+\frac{1}{2}, j+
\frac{1}{2}}(t)|\leq{e}^{M t}\sum_{i=1}^{L-1}
\sum_{j=0}^i|n_{i+\frac{1}{2}, j+\frac{1}{2}}(0)|,
\end{equation}
where $\tilde{B}=2\|\tilde{\beta}\|_{\infty}, M=2B-b,
B=\|\beta\|_{\infty}$, and $b=\min\limits_{t}\min\limits_{i,
  j}\beta_{i+\frac{1}{2}, j+\frac{1}{2}}(t)$. The
$\textbf{L}^{\infty}$ bound is given by $\| n^h(t)\|_{\infty}\leq
\max\{\frac{1}{g_{\rm min}}{\tilde{B}}e^{MT}\|n(0)\|_1,
\|n^h(0)\|_{\infty}\}e^{2\tilde{g}'t}$ where $\tilde{g}'$ is the
$\textbf{L}^{\infty}$ bound of $\partial g/\partial x, \partial
g/\partial y$.

\vspace{2mm}

\textbf{Proof} For the summation of $n$ over all grid points, we
multiply the first equation in \ref{DisForm} by
sign$(n_{i+\frac{1}{2}, j+\frac{1}{2}})$ for each $i, j\leq i$,

\begin{equation}
\begin{aligned}
h^2\frac{\dd}{\dd t}|n_{i+\frac{1}{2}, j+\frac{1}{2}}(t)|+
hg_{i+1, j+\frac{1}{2}}(t)|n_{i+\frac{1}{2}, j+\frac{1}{2}}(t)|+
hg_{i+\frac{1}{2}, j+1}(t)|n_{i+\frac{1}{2}, j+\frac{1}{2}}(t)|+
h^2\beta_{i+\frac{1}{2}, j+\frac{1}{2}}(t)|n_{i+\frac{1}{2}, j+\frac{1}{2}}(t)|\leq\\
hg_{i, j+\frac{1}{2}}(t)|n_{i-\frac{1}{2}, j+\frac{1}{2}}(t)|+
hg_{i+\frac{1}{2}, j}(t)|n_{i+\frac{1}{2}, j-\frac{1}{2}}(t)|
\end{aligned}
\end{equation}
By multiplying the second equation in \ref{DisForm} by
sign$(n_{i+\frac{1}{2}, j+\frac{1}{2}})$ for each $i, j\leq i$ pair
and summing over index $\sum_{i=1}^{L-1}\sum_{j=0}^{i-1}$,
\begin{equation*}
\begin{aligned}
h^2\sum_{i=1}^{L-1}\sum_{j=0}^{i-1}|n_{i+\frac{1}{2},
  j+\frac{1}{2}}(t)|
%
+h\sum_{j=0}^{i-1}g_{L,
  j+\frac{1}{2}}(t)|n_{L-1+\frac{1}{2},
  j+\frac{1}{2}}(t)|\hspace{2cm} \\ + h^2\sum_{i=1}^{L-1}\sum_{j=0}^{i-1}\beta_{i+\frac{1}{2},
  j+\frac{1}{2}}(t)|n_{i+\frac{1}{2}, j+\frac{1}{2}}(t)|
\leq{h}\sum_{i=0}^{L-1}g_{i+\frac{1}{2}, 0}(t)|n_{i+\frac{1}{2},
  -\frac{1}{2}}(t)|.
\end{aligned}
\end{equation*}

\noindent We can simplify the above expression to

\begin{eqnarray*}
h^2\frac{d}{dt}\sum_{i=1}^{L-1}\sum_{j=0}^{i-1}
\vert n_{i+\frac{1}{2}, j+\frac{1}{2}}(t)\vert & + & h^2\sum_{i=1}^{L-1}\sum_{j=0}^{i-1}
\beta_{i+\frac{1}{2}, j+\frac{1}{2}}\vert n_{i+\frac{1}{2}, j+\frac{1}{2}}(t)\vert \\
\: & \leq & 2h^3\sum_{i=0}^{L-1}\big| \sum_{\ell=i}^{L-1}\sum_{j=0}^{\ell-1}
\tilde{\beta}_{\ell+\frac{1}{2}, j+\frac{1}{2}}((i+1/2)h,t)
n_{\ell+\frac{1}{2}, j+\frac{1}{2}}(t)\big| \\
\: &  \leq &  2h^2\sum_{\ell=1}^{L-1}\sum_{j=0}^{\ell-1}
|\beta_{\ell+\frac{1}{2}, j+\frac{1}{2}}(t)||n_{\ell+\frac{1}{2}, j+\frac{1}{2}}(t)|.
\end{eqnarray*}
We then have

\begin{equation*}
\frac{\dd}{\dd t}\sum_{i=1}^{L-1}\sum_{j=0}^{i-1}
|n_{i+\frac{1}{2}, j+\frac{1}{2}}(t)|\leq({2B-b})
\sum_{i=1}^{L-1}\sum_{j=0}^{i-1}|n_{i+\frac{1}{2}, j+\frac{1}{2}}(t)|,
\end{equation*}
which yields
\begin{equation}
\sum_{i=1}^{L-1}\sum_{j=0}^{i-1}|n_{i+\frac{1}{2}, j+\frac{1}{2}}(t)|
\leq{e}^{Mt}\sum_{i=1}^{L-1}\sum_{j=0}^{i-1}
|n_{i+\frac{1}{2}, j+\frac{1}{2}}(0)|.
\label{L1NORM}
\end{equation}
\ref{L1NORM} states that the $l^1$ norm of all the values on the grid
points are uniformly bounded and independent of $h$. Next,
we estimate the $\textbf{L}^{\infty}$ bound of $n^h$. First, we
consider $j=0$ and assume $S(t)=\max\limits_{1\leq{i}\leq{L-1}}
|n_{i+\frac{1}{2},\frac{1}{2}}(t)|e^{-\tilde{g}'t}$ for $t\in[0,
  T]$. For the maximum value of $S$ at some index $i$, we find
\small{\begin{equation}
\begin{array}{l}
h^{2}\frac{\dd|n_{i+\frac{1}{2}, \frac{1}{2}}(t)|}{\dd t}+h(g_{i+1,
  \frac{1}{2}}(t)|n_{i+\frac{1}{2}, \frac{1}{2}}(t)|-g_{i,
  \frac{1}{2}}(t)|n_{i-\frac{1}{2}, \frac{1}{2}}(t)|) + h(g_{i+\frac{1}{2}, 1}(t)|n_{i+\frac{1}{2},
  \frac{1}{2}}(t)|-g_{i+\frac{1}{2}, 0}(t)|n_{i+\frac{1}{2},
  -\frac{1}{2}}(t)|)\leq 0, \nonumber \\[13pt] 
h^2\frac{\dd|n_{i+\frac{1}{2},
    \frac{1}{2}}(t)|}{\dd t}+hg_{i+1, \frac{1}{2}}(t)|n_{i+\frac{1}{2},
  \frac{1}{2}}(t)|-g_{i+\frac{1}{2}, 0}(t)|n_{i+\frac{1}{2},
  -\frac{1}{2}}(t)| \leq 0,\quad  i=1.
\end{array}
\end{equation}}
\normalsize
and
\begin{equation*}
\frac{\dd(|n_{i+\frac{1}{2},
    \frac{1}{2}}(t)|e^{-\tilde{g}'t})}{\dd t}+h^{-1}g_{i+\frac{1}{2},
    1}(t)|n_{i+\frac{1}{2},
  \frac{1}{2}}(t)|e^{-\tilde{g}'t}\leq h^{-1}g_{i+\frac{1}{2},
    0}(t)|n_{i+\frac{1}{2}, -\frac{1}{2}}(t)|e^{-\tilde{g}'t},
\end{equation*}
By the assumption that $g(x, y,t)\geq{g}_{\rm min}(t)\geq{g}_{\rm min}>0$ and
$g< K (T+1+\Omega(T))$, we have

\begin{equation}
\frac{\dd(|n_{i+\frac{1}{2},
    \frac{1}{2}}(t)|e^{-\tilde{g}'t})}{\dd t}+ h^{-1}g_{\rm min}(t) |n_{i+\frac{1}{2},
  \frac{1}{2}}(t)|e^{-\tilde{g}'t}
\leq h^{-1}\left(\frac{g_{\rm min}(t)}{g_{\rm min}}\right)\max_{1\leq{i}\leq{L-1}}|g_{i+\frac{1}{2},
  0}(t)n_{i+\frac{1}{2}, -\frac{1}{2}}(t)|.
\end{equation}
%
%
Finally, defining $G(t)=h^{-1}\int_0^t g_{\rm min}(s) \dd{s}$ yields

\begin{equation*}
\frac{\dd(|n_{i+\frac{1}{2},
    \frac{1}{2}}(t)|e^{-\tilde{g}'t}e^{G(t)})}{\dd t}
\leq {1\over h}\left(\frac{g_{\rm min}(t)}{g_{\rm min}}\right)
\max_{1\leq{i}\leq{L-1}}
|g_{i+\frac{1}{2},0}(t)n_{i+\frac{1}{2}, -\frac{1}{2}}(t)|e^{G(t)}.
\end{equation*}
From the $\textbf{L}^{1}$ bound, we can deduce

\begin{equation*}
\max\limits_{t}\max_{1\leq{i}\leq{L-1}}|g_{i+\frac{1}{2}}(t)n_{i+\frac{1}{2},
  -\frac{1}{2}}(t)|\leq{h^2}{\tilde{B}}e^{MT}\|n^h(0)\|_1
\leq{\tilde{B}e^{MT}\|n(0)\|_1},\quad 
t>0
\end{equation*}
and conclude that for the function $S(t)e^{G(t)}$
\begin{equation}
S(t)e^{G(t)}\leq S(0) + \frac{1}{g_{\rm min}}\tilde{B}e^{MT}\|n(0)\|_1(e^{G(t)}-1),
\end{equation}
and $S(t)\leq\max\limits_{1\leq{i}\leq{L-1}}\{n_{i+\frac{1}{2},\frac{1}{2}}(0),\frac{1}{g_{\rm
    min}} {\tilde{B}}e^{MT}\|n(0)\|_1\}$, which then gives the
$\textbf{L}^{\infty}$ bound for the pointwise solution $n^h$ when
$j=0$.

Now, we  estimate $|n_{i+\frac{1}{2}, j+\frac{1}{2}}(t)|$
by first defining  $P(t)\equiv \max\limits_{0\leq{i}\leq{L-1},
    0\leq{j}\leq{i-1}}\{|n_{i+\frac{1}{2},
    j+\frac{1}{2}}(t)|e^{-2\tilde{g}'t}\}$. At a fixed time $t$,
specific values of $i$ and $j$ define $P(t)$. If the maximum occurs 
at $j=0$, $P(t) = S(t)e^{-\tilde{g}'t}$. If the maximum 
occurs at $i-1 > j > 0$, we have


\begin{eqnarray}
h\frac{\dd}{\dd t}(|n_{i+\frac{1}{2},
  j+\frac{1}{2}}(t)|e^{-2\tilde{g}'t}) = - \left[
g_{i,j+\frac{1}{2}}(t)|n_{i+\frac{1}{2}, j+\frac{1}{2}}(t)| - g_{i+1,
    j+\frac{1}{2}}(t)|n_{i+\frac{1}{2}, j+\frac{1}{2}}(t)|\right. \hspace{1cm} \nonumber \\
\left. +g_{i+\frac{1}{2}, j}(t)|n_{i+\frac{1}{2}, j+\frac{1}{2}}(t)| -
  g_{i+\frac{1}{2}, j+1}(t)|n_{i+\frac{1}{2},
    j+\frac{1}{2}}(t)| + 2h\tilde{g}'|n_{i+\frac{1}{2},j+\frac{1}{2}}(t)|\right] 
e^{-2\tilde{g}'t}\leq 0,
\label{P1}
\end{eqnarray}
while if the maximum occurs at $j=i-1>0$, we have

\begin{equation}
\frac{\dd}{\dd t}(|n_{i+\frac{1}{2},
  j+\frac{1}{2}}(t)|e^{-2\tilde{g}'t})\leq \left[h^{-1}\left(g_{i+\frac{1}{2},
    j}(t) - g_{i+1, j+\frac{1}{2}}(t)\right)|n_{i+\frac{1}{2},
  \frac{1}{2}}(t)|-2\tilde{g}'|n_{i+\frac{1}{2},
  j+\frac{1}{2}}(t)|\right] e^{-2\tilde{g}'t}\leq 0.
\label{P2}
\end{equation}
In Eqs.~\ref{P1} and \ref{P2}, $i, j$ are the maximizing indices 
that define $P(t)$.

For any $t\in(0, T]$ we can find a minimum $\tilde{t}<t$ such that
$P(v)>S(v)e^{-\tilde{g}'v}$ for $v\in(\tilde{t}, t]$.  If
$\tilde{t}=0$, and since $P(t)$ is nonincreasing from Eq.~\ref{P2},
$P(t)\leq P(0)= \|n^h(0)\|_{\infty}$. If $t>\tilde{t} >0$,
$P(t)\leq{P(\tilde{t})}\leq{S(\tilde{t})}\leq\max\limits_{0\leq{t}\leq{T}}{S(t)}$,
while if $\tilde{t}=t$,
$P(t)=S(t)\leq\max\limits_{0\leq{t}\leq{T}}{S(t)}$. Thus, 
$P(t) = \| n^{h}(t)\|_{\infty}
e^{-2\tilde{g}'t}\leq\max\{\max\limits_{0\leq{t}\leq{T}}\{S(t)\},
\|n^h(0)\|_{\infty}\}$ and

\begin{equation}
\|n^h(t)\|_{\infty}\leq \max\{\max\limits_{0\leq{t}\leq{T}}\{S(t)\},
\|n^h(0)\|_{\infty}\}e^{2\tilde{g}'t},
\end{equation}
giving the second conclusion in Theorem A.2 that the
  $\textbf{L}^{\infty}$ bound is uniform and independent of $h$.

\subsubsection{Existence of the weak solution}

For a given time $T<\infty$, we can take the grid size
$h(k)=\Omega(T)/k \to 0$ by letting the integer $k \to
\infty$. Spatially piecewise constant functions can then be defined
based on the sequence of vector functions $\{n^{h(k)}\}$.  By setting
$n^h_{i+\frac{1}{2}, i+\frac{1}{2}}(t)=0$, we define $n^h(x,y,t)$,
$\beta^h$, and $\tilde{\beta}^h$ as

\begin{align*}
n^h(x, y, t) = & \sum_{i=0}^{k-1}\sum_{j=0}^{i-1}n^h_{i+\frac{1}{2}, j+\frac{1}{2}}(t)
\mathds{1}(ih\leq{x}<(i+1)h, jh\leq{y}<(j+1)h), \nonumber \\
\beta^h(x, y, t)= & \sum_{i=0}^{k-1}\sum_{j=0}^{i}
\beta_{i+\frac{1}{2}, j+\frac{1}{2}}(t)\mathds{1}(ih\leq{x}<(i+1)h, jh\leq{y}<(j+1)h), \nonumber \\
\tilde{\beta}^h(x, y, z, t) = & \sum_{i=0}^{k-1}\sum_{j=0}^{i-1}\sum_{\ell=0}^{i-1}
\tilde{\beta}_{i+\frac{1}{2}, j+\frac{1}{2}}((\ell +\frac{1}{2})h, t)
\mathds{1}(ih\leq{x}<(i+1)h, jh\leq{y}<(j+1)h, \ell h\leq{z}<(\ell +1)h), \nonumber \\
n^h(x, 0, t)= &  n^h_{i+\frac{1}{2}, -\frac{1}{2}}(t), \quad ih\leq x<(i+1)h,
\end{align*}

\noindent where above, $h = h(k)$ and $\mathds{1}$ is the indicator function. 
Since there is an upper bound for both $\beta$ and $\tilde{\beta}$,
and both $\beta, \tilde{\beta}$ are continuous, we have the following
result

\begin{align*}
\lim\limits_{k\rightarrow\infty}\beta^{h(k)}(x, y, t)
\rightarrow & \beta(x, y, t) \; a.e. \quad 0\leq\beta^{h(k)}\leq{\|\beta\|_{\infty}<\infty},\\
\lim\limits_{k\rightarrow\infty}\tilde{\beta}^{h(k)}(x, y, z, t)
\rightarrow & \beta(x, y, z, t) \; a.e. \quad 
0\leq\tilde{\beta}^{h(k)}\leq\|\tilde{\beta}\|_{\infty}<\infty,\\
\lim\limits_{k\rightarrow\infty}n^{h(k)}(x, y, 0)\rightarrow & {n(x, y, 0)}\; a.e..
\end{align*}
Then, we can apply Theorem A.2 to the 
piecewise constant solutions $n^{h(k)}$ of Eqs.~\ref{DisForm}.

\textbf{Corollary A.3} Under the conditions of Theorem A.2,
for any $t\in[0, T]$ and any $h$,
\begin{equation}
\int_0^{\Omega(T)}\!\!\dd y \int_0^{\Omega(T)}\!\!\dd x\, |n^h(x, y,
t)| \leq{e}^{Mt}\int_0^{\Omega(0)}\!\!\dd y\int_0^{\Omega(0)}\!\!\dd
x\, |n^h(x, y, 0)|
\end{equation}
and 
\begin{equation}
\|n^h(t)\|_{\infty}\leq \max \{\vert n(0)\vert_{\infty},
   {B}e^{MT}\vert n(0)\vert_{1}\}e^{2\tilde{g'}t},
\end{equation}
where $B, M, \tilde{g}'$ are defined in Theorem A.2. 
The proof is the direct consequence of Theorem A.2.

The sequence of piecewise constant functions $\{n^{h(k)}\}$ is
uniformly bounded and
$n^{h(k)}\in\textbf{L}^1\cap\textbf{L}^{\infty}([0,\Omega(T)]^{2}\cap\{y<x\}\times
[0,T))$, so $n^{h(k)}$ are all $\textbf{L}^2$ functions.  There exists
  a function $n\in\textbf{L}^{2}([0,\Omega(T)]^{2}\cap\{y<x\}\times
  [0,T))$ and a subsequence $k_i\rightarrow\infty$ that satisfies
    $n^{h(k_i)}\rightharpoonup{n}$.  Since
    $\textbf{L}^2([0,\Omega(T)]^{2}\cap\{y<x\}\times [0,T))$ implies
      $\textbf{L}^1$ integrability, we can deduce that $n$ is an
      $\textbf{L}^1$ function as desired.

To prove $n^{h(k_i)}\rightharpoonup{n}$, we need only to verify that
there exists a subsequence $n^{h(k_i)}$ such that for all test
functions $f\in\textbf{L}^2$, $\int_0^T\!\!\dd
t\int_0^{\Omega(T)}\!\!\dd x \int_0^x\!\!\dd y \, n^{h(k_i)}f
\rightarrow\int_0^T\!\!\dd t\int_0^{\Omega(T)}\!\!\dd x\int_0^x\!\!\dd
y\, nf$. Since $\textbf{L}^2$ space is separable, we have a countable
set of basis function $\{b_i(x, y, t)\}$ for the space
$\textbf{L}^{2}([0,\Omega(T)]^{2}\cap\{y<x\}\times [0,T))$.  Thus,
  every $n^{h(k)}$ can be decomposed as $n^{h(k)} =
  \sum_{i=1}^{\infty}\alpha^k_ib_i$.  The sequence $\{n^{h(k)}\}$ is
  uniformly $\textbf{L}^{\infty}$ bounded, so $\sum\alpha_k^2$ are all
  uniformly bounded. We can then select a subsequence $\{n^{h(k_i)}\}$
  from $\{n^{h(k)}\}$ satisfying
  $\lim_{i\rightarrow\infty}\alpha^{k_i}_j=\alpha_j$ so that
  $\sum_{i=1}^{\infty}\alpha_j^2 <\infty$. If we set
  $n=\sum_{i=1}^{\infty}\alpha_i b_i$, then, by decomposing any test
  function $\Psi\in\textbf{L}^2([0,\Omega(T)]^{2}\cap\{y<x\}\times
  [0,T))$ by $\Psi = \sum_{i=1}^{\infty}\gamma_i b_i$, we have

\begin{equation}
\lim_{i\rightarrow\infty}\bigg| \int_0^T\!\!\dd t \int_0^{\Omega(T)}\!\!\dd x 
\int_0^x\!\!\dd y \, \left(n^{h(k_i)}-n\right)
\Psi\bigg|
= \bigg| \sum_{s=1}^{\infty}(\alpha^{k_i}_s-\alpha_s)\gamma_s\bigg|=0,
\end{equation}
which gives the result $n^{h(k_i)}\rightharpoonup{n}$.  

We can show that $n$ is a weak solution by multiplying the
  first two of Eqs.~\ref{DisForm} by a test function
  $\Psi\in\textbf{C}^1([0,\Omega(T)]^{2}\times [0,T))$,
    $\Psi(x,y,T)=0, \Psi(x,y,t)=0, y\geq{x}$ for which

\begin{equation*}
\Psi_{i+\frac{1}{2},
  j+\frac{1}{2}}(t)\equiv \frac{1}{h^2}\int_{x_i}^{x_{i+1}}\!\!\dd
x\int_{y_j}^{y_{j+1}}\!\!\dd y \, \Psi(x,y,t), \quad j\leq{i}.
\end{equation*}
For a given $L\in\textbf{N}^+$ and $h=\frac{\Omega(T)}{L}$,

{\small\begin{equation*}
\begin{aligned}
\int_0^T\!\!\dd t \sum_{i=1}^{L-1}\sum_{j=0}^{i-1}\bigg(
h^2\frac{\dd n^h_{i+\frac{1}{2},
    j+\frac{1}{2}}(t)}{\dd t}\Psi_{i+\frac{1}{2}, j+\frac{1}{2}}(t) +
h\left[g_{i+1, j+\frac{1}{2}}(t)n^h_{i+\frac{1}{2}, j+\frac{1}{2}}(t)-g_{i,j+
\frac{1}{2}}(t)n^h_{i-\frac{1}{2},
    j+\frac{1}{2}}\right]\Psi_{i+\frac{1}{2}, j+\frac{1}{2}}(t) \\
+ h[g_{i+\frac{1}{2}, j+1}(t)n^h_{i+\frac{1}{2},
    j+\frac{1}{2}}(t)-g_{i+\frac{1}{2}, j}(t)n^h_{i+\frac{1}{2},
    j-\frac{1}{2}}]\Psi_{i+\frac{1}{2}, j+\frac{1}{2}}(t)+
h^2\beta_{i+\frac{1}{2}, j+\frac{1}{2}}(t)n^h_{i+\frac{1}{2},
  j+\frac{1}{2}}\Psi_{i+\frac{1}{2}, j+\frac{1}{2}}(t)\bigg) 
\\
= \int_0^T\!\!\dd t \sum_{i=1}^{L-1}h
g_{i+\frac{1}{2}, i}(t)n^h_{i+\frac{1}{2}, i-\frac{1}{2}}(t)\Psi_{i+\f{1}{2}, i-\f{1}{2}}(t), 
\,\,\, n^{h}_{i+\frac{1}{2}, i+\frac{1}{2}} = 0.
\end{aligned}
\end{equation*}}
Integrating the above equation by parts with respect to time, we find

{\small\begin{equation}
\begin{aligned}
\int_0^T\!\!\dd t \left[
\sum_{i=1}^{L-1}\sum_{j=0}^{i-1} h^2n^h_{i+\frac{1}{2},
  j+\frac{1}{2}}(t)\frac{\dd\Psi_{i+\frac{1}{2},
    j+\frac{1}{2}}(t)}{\dd{t}} + h\sum_{i=1}^{L-2}\sum_{j=0}^{i-1} 
g_{i+1,j+\frac{1}{2}}(t)n^h_{i+\frac{1}{2},j+\frac{1}{2}}(t)
(\Psi_{i+\frac{3}{2}, j+\frac{1}{2}}(t)-\Psi_{i+\frac{1}{2}, j+\frac{1}{2}}(t))\right. \\
\left. + h\sum_{i=1}^{L-1}\sum_{j=0}^{i-2} g_{i+\frac{1}{2},
  j+1}(t)n^h_{i+\frac{1}{2}, j+\frac{1}{2}}(t)(\Psi_{i+\frac{1}{2},
  j+\frac{3}{2}}(t)-\Psi_{i+\frac{1}{2}, j+\frac{1}{2}}(t))\right] 
 =\hspace{1.7cm} \\
-h^2\sum_{i=0}^{L-1}\sum_{j=0}^{i-1}n^h_{i+\frac{1}{2},
  j+\frac{1}{2}}(0)\Psi_{i+\frac{1}{2}, j+\frac{1}{2}}(0) 
- h\int_0^{T}\!\!\dd t \sum_{i=1}^{L-1}g_{i+\frac{1}{2},
  0}(t)n^h_{i+\frac{1}{2}, -\frac{1}{2}}(t)\Psi_{i+\frac{1}{2},
  \frac{1}{2}}(t)\hspace{2cm} \\
-\int_0^T\!\!\dd t \sum_{i=1}^{L-1}h
g_{i+\frac{1}{2}, i}(t)n^h_{i+\frac{1}{2}, i-\frac{1}{2}}(t)\Psi_{i+\f{1}{2}, i-\f{1}{2}}(t)\hspace{3cm} \\
+h\int_0^T\!\!\dd t \left[\sum_{j=0}^{L-2}g_{L, j+\frac{1}{2}}(t)n^h_{L-\frac{1}{2},
  j+\frac{1}{2}}(t)\Psi_{L-\frac{1}{2},
  j+\frac{1}{2}}(t) + \sum_{i=1}^{L-1}\sum_{j=0}^{i-1}h^2\beta_{i+\frac{1}{2},
  j+\frac{1}{2}}(t)n^h_{i+\frac{1}{2},
  j+\frac{1}{2}}(t)\Psi_{i+\frac{1}{2}, j+\frac{1}{2}}(t)\right] .
  \label{Inte Exi}
\end{aligned}
\end{equation}}
Since $\Psi_{i+\frac{3}{2}, j+\frac{1}{2}}(t)-\Psi_{i+\frac{1}{2},
  j+\frac{1}{2}}(t) = \int_{ih}^{(i+1)h}\!\!\dd
x\int_{jh}^{(j+1)h}\!\!\dd y \int_{x}^{x+h}\!\!\dd s\,
\frac{\partial\Psi}{\partial{s}}(s,y,t)$, $|n^h|$ is uniformly bounded
while $g$ is $\textbf{C}^1$ continuous. From above we can pick a
\added{subsequence} in $\{n^{h(k)}\}$, \added{denoted by
  $n^{h(k_i)}\rightharpoonup{n}$.} We use
$n^{h}=n^{h(k_i)}$ in the above formula. Since $\Psi\in\textbf{C}^1[0,
  T]\times[0, \Omega(T)]^2$, given any $\Psi$ we have a positive upper
bound $R(\Psi)<\infty$ for $\Psi$ and any of its first
derivatives. Thus,

\begin{equation*}
\begin{aligned}
\bigg|\int_0^T\!\!\dd t \sum_{i=1}^{L-1}
\sum_{j=0}^{i-1}\left(h^2n^{h(k_i)}_{i+\frac{1}{2},
j+\frac{1}{2}}(t)\frac{\dd\Psi_{i+\frac{1}{2},
j+\frac{1}{2}}(t)}{\dd t}\right) -
\int_0^T\!\!\dd t \int_0^{\Omega(T)}\!\!\dd x \int_0^{x}\!\!\dd y\, 
n^{h(k_i)}(x, y, t)\frac{\partial\Psi(x,y,t)}{\partial{t}}\bigg| 
\leq \\ \int_0^T\!\!\dd t \sum_{i=0}^{L-1}
\int_{ih}^{(i+1)h}\!\!\dd x \int_{ih}^x\!\!\dd y\, \bigg|n^{h(k_i)}(
x, y, t)\frac{\partial\Psi(x, y, t)}{\partial{t}}\bigg|.
\end{aligned}
\end{equation*}
As $h\to 0$, $|\int_0^T\!\!\dd
t\sum_{i=0}^{L-1}\int_{ih}^{(i+1)h}\!\dd x \int_{ih}^x\!\dd y\,
n^{h(k_i)}(x,y, t)\frac{\partial\Psi(x, y, t)}{\partial{t}}| \to 0$
since $\frac{\partial\Psi}{\partial{t}}$ and $n^{h(k_i)}$ are all
bounded. Moreover,

\begin{equation}
\int_0^T\!\!\dd t \int_0^{\Omega(T)}\!\!\dd x\int_0^{x}\!\!\dd y \,
h^2n^{h(k_i)}(x,y,t)\frac{\partial\Psi(x,y,t)}{\partial{t}}
\rightarrow \int_0^T\!\!\dd t\int_0^{\Omega(T)}\!\!\dd x\int_0^{x}\!\!\dd y\, 
h^2n(x, y, t)\frac{\partial\Psi(x, y, t)}{\partial{t}}
\label{conv}
\end{equation}
so that the first term in Eq.~\ref{Inte Exi} tends to the limit in
Eq.~\ref{conv}. By the same procedure and using the condition that $g$
is uniformly continuous in $[0, T]\times[0, \Omega(t)]^2$($g$ is
$\textbf{C}^1$), it is easy to verify that the second and third terms
on the LHS of Eq.~\ref{Inte Exi} tend to $\int_0^T\!\!\dd t
\int_0^{\Omega(T)}\!\!\dd x\int_0^{x}\!\!\dd y\,
(gn)(x,y,t)\frac{\partial{\Psi}}{\partial{x}}$ and $\int_0^T\!\!\dd
t\int_0^{\Omega(T)}\!\!\dd x\int_0^{x}\!\!\dd y\, (gn)(x,
y,t)\frac{\partial{\Psi}}{\partial{y}}$, respectively.

It is also easy to verify that the first and second terms 
on the RHS of \ref{Inte Exi}
%
%
tend to \\ 
$-\int_0^T\!\!\dd t\int_0^{\Omega(T)}\!\!\dd
x\int_0^x\!\!\dd y\, n(x,y,0)\Psi(x,y,0)$ and $-2\int_0^T\!\!\dd t
\int_0^{\infty}\!\!\dd x\int_0^x\!\!\dd y\int_0^x\!\!\dd z\,
\Psi(z,0,t)\tilde{\beta}(x,y,z,t)n(x,y,t)$, respectively.  The third
term on the RHS of \ref{Inte Exi} $h\int_{0}^{T}\!\dd t
\sum_{i=1}^{L-1} g_{i+\frac{1}{2},i}(t)
n^{h}_{i+\frac{1}{2},i-\frac{1}{2}}(t)\Psi_{i+\frac{1}{2},i-\frac{1}{2}}(t)$
tends to 0 since $\Psi$ is $\textbf{C}^1$ continuous and is 0 on the
boundary $x=y$. Since $\Psi$ is continuous and is 0 at $x=\Omega(T)$,
\\ 
$h\int_0^T\sum_{j=0}^{L-2}g_{L,
  j+\frac{1}{2}}(t)n^h_{L-\frac{1}{2},j+\frac{1}{2}}(t)
\Psi_{L-\frac{1}{2},j+\frac{1}{2}}(t)\dd t \rightarrow 0$ as
$h\rightarrow 0$. Finally, the last term on the RHS of of \ref{Inte
  Exi}
$h\int_0^T\sum_{i=1}^{L-1}\sum_{j=0}^{i-1}h^2\beta_{i+\frac{1}{2},
  j+\frac{1}{2}}(t)n^h_{i+\frac{1}{2},
  j+\frac{1}{2}}(t)\Psi_{i+\frac{1}{2}, j+\frac{1}{2}}(t)\dd{t}$
$\rightarrow \int_0^T\!\dd t \int_0^{\Omega(T)}\!\!\dd
x\int_0^{x}\!\dd y\, \beta(x,y,t)n(x, y, t)\Psi(x,y,t)$.

By passing to the limit $h\rightarrow 0$, we conclude that $n$
exactly satisfies the condition of a weak solution in
\ref{AdjPro}. Since the numerical solution obtained by the scheme in
Appendix B is a discretization in time for the ODE system
\ref{DisForm} it is an approximation to the solution of
\ref{DisForm}. Provided $h,\Delta t \to 0$ 
satisfies the CFL condition $2\|g\|_{\infty}\Delta t < h$,
conclude that at least a subsequence of the numerical
solutions converge to the unique weak solution of
\ref{PDEForm}.
%
%
Furthermore, recently, the existence to an eigenpair of the
adder-sizer PDE \ref{PDEForm} under specific smooth conditions
satisfied by the coefficients $g, \beta, \tilde{\beta}$ has been
proved in \cite{GABRIEL2019}, allowing for studying asymptotic
behavior of the solution.

\section{Numerical Scheme}

We denote $\u(t) = \{\n_{1}(t), \n_{2}(t), \ldots,
\n_{L-1}(t)\}^{\texttt{T}}$ where $\n_{j}(t) =
\{n_{\frac{1}{2},j-\frac{1}{2}}, n_{1+\frac{1}{2},j-\frac{1}{2}},
\ldots, n_{L-\frac{1}{2},j-\frac{1}{2}}\}$ and $n_{i\leq j} = 0$.
Equations \ref{NumSch} and \ref{NumSch2} can then be written in the
form $\u(t+\Delta t) = {\bf A}(t)\u(t)$, where

\begin{equation}
{\bf A}(t) = \left[\begin{array}{cccccccc}
{\bf B}_{1}+{\bf C}_{1} & {\bf C}_{2} & {\bf C}_{3} & {\bf C}_{4} & \cdots  & {\bf C}_{L-2} & {\bf C}_{L-1}  \\
{\bf D}_{2} & {\bf B}_{2} & 0 & 0 & \cdots & 0 & 0 \\
0 & {\bf D}_{3} & {\bf B}_{3} & 0 & \cdots & 0 & 0 \\
\vdots & \vdots & \vdots & \vdots & \vdots & \vdots & \vdots  \\
0 & 0 & 0 & 0 &  \cdots & {\bf B}_{L-2} & 0 \\ 
0 & 0 & 0 & 0 &  \cdots & {\bf D}_{L-1} & {\bf B}_{L-1}  \\
\end{array}\right],
\end{equation}
is made up of the following $L-1$ $L\times L$ matrices

\begingroup
\renewcommand*{\arraystretch}{1.4}
\begin{equation*}
{\bf B}_i=
\left[
    \begin{array}{r@{}c|c@{}l}
 & & & \\[-10pt]
 & \mbox{\large 0 $(i\times i)$} & \mbox{\large 0 $(i\times (L-i))$} & \\[-10pt]
 & & & \\\hline
 & & & \\[-10pt] 
 & \mbox{\large 0 $((L-i)\times i)$} &  
      {\bf b}_{i} & \\[-10pt]
& & & 
    \end{array} \right],\quad 
{\bf C}_i=
\left[
    \begin{array}{r@{}c|c@{}l}
 & & & \\[-8pt]
      &  \mbox{\large 0 $(1\times i)$} & \mbox{\large 0 $(1\times (L-i))$} & \\[-8pt]
      & & & \\\hline
       & & & \\[-6pt]
       &  \mbox{\large 0 $((L-1)\times i)$} & {\bf c}_{i} &  \\[-6pt]
      & &  & \end{array}\right],
\end{equation*}
\endgroup

\begingroup
\renewcommand*{\arraystretch}{1.4}
\begin{equation*}
\mbox{and} \quad {\bf D}_i=
\left[
    \begin{array}{r@{}c|c@{}l}
 & & & \\[-10pt]
 &  \mbox{\large 0 $(i\times i)$} & \mbox{\large 0 $(i\times (L-i))$} & \\[-10pt]
 & & & \\\hline
 & & & \\[-10pt]
 & \mbox{\large 0 $((L-i)\times i)$} & {\bf d}_{i} & \\[-10pt]
 & & &  \end{array}\right],
\end{equation*}
\endgroup
in which ${\bf b}_{i}$ is a lower bidiagonal matrix with diagonal 

\begin{equation}
  {\rm diag}({\bf b}_{i})  = 1-{1\over h}g_{j+1,i-{1\over 2}}(t)\dd t
  -{1\over h}g_{j+\frac{1}{2},i}(t)\dd t - 
\beta_{j+{1\over 2},i-{1\over 2}}(t)\dd t, \quad j=i,i+1,\ldots, L-1,
\end{equation}
and lower off diagonal $({\bf b}_{i})_{-1} = g_{j,i-{1\over
    2}}(t){\dd t \over h},\quad j=i+1,\ldots, L-1,$


\begin{equation}
({\bf c}_i)_{sj} = \left\{
\begin{array}{ll}
\tilde{\beta}_{i-\f{1}{2}+j,i-\f{1}{2}}
((s+\f{1}{2})h, t)\dd{t}, \qquad &   i+j-s-1>0,\, i+j \leq L\\
0 &  {\rm otherwise.}
\end{array}
\right.
\end{equation}

%
\noindent and ${\bf d}_{i}$ is a diagonal matrix
${\rm diag}({\bf d}_{i}) = g_{j+{1\over 2}, i-1}(t) {\dd t \over h}, \quad
  j=i, i+1, \ldots, L-1.$

\section{Monte-Carlo Simulations}
In this section we describe the implementation of our Monte-Carlo
simulations of the process underlying the adder-sizer mechanism.
Suppose we have a list of cells at time $t$ denoted by
$S(t)=\{c_1(x_i, y_i, t, b_1), ..., c_i(x_i, y_i, t, b_i)\}$,
where $x_i$ is cell $c_i$'s volume and $y_i$ is its added volume. The
cell's division factor $b_i$ is determined at birth, which is drawn
from a uniform distribution $\textbf{U}(0, 1)$.

Suppose we have a $\beta$ of the form \ref{BETA_GAMMA} and
$\tilde{\beta}$ of the form \ref{TildeBeta}. We set the maximum
  allowable time step to $\Delta{t}=0.01$ and determine the next state
of the system at time $t'$ by the following

\vspace{0.05in}
\begin{itemize}
\item Step 1: For each cell $i$, calculate its age $a_i$ at time $t$
  by the exponential growth law
  $\frac{\textrm{d}x}{\textrm{d}t}=\lambda x$.  We require that
  $G_i=\int_{0}^{a_i}\gamma(a')\dd{a'} < b_i$ at the beginning of each
  step for every $i$.
\item Step 2: For each cell, calculate
  $G_i=\int_{0}^{a_i+\Delta{t}}\gamma(a')\dd{a'}$. If $G_i\geq b_i$,
  then we numerical calculate a $\Delta{t}_i$ such that
  $\int_{0}^{a_i+\Delta{t}_i}\gamma(a')\dd{a'}\approx{b_i}$.
\item Step 3: Choose the smallest $\Delta{t}_i$ among all possible
  $\Delta{t}_i$s as the new time step, set time $t'=t+\Delta{t}_i$ and
  let all cells gain an extra volume $\lambda{x_i}\Delta{t}_i$. If
  there is no such $\Delta{t}_i$, which means $G_i<b_i$ for every $i$,
  go to step 5.
\item Step 4: Remove cell $i$ from $S(t')$, record its volume $x$ at
  $t'$, and generate the random numbers $r$ from the distribution
  $h(r)$ and $b^{m}, b^{m+1}$ from ${\bf U}(0,1)$. Then, add two new
  cells in $S(t')$ labeled by $c_{m}(rx,0,t,b^{m})$ and
  $c_{m+1}(x-rx,0,t,b^{m+1})$.
\item Step 5: If $G_i<b_i$ for all $i$, set $t=t'$ and let all
  cells gain an extra volume $\lambda{x_i}\Delta{t}_i$.
\item Step 6: Return to step 1 until $t'>t_{\rm max}$, the maximum
  time of the simulation.
\end{itemize}
\vspace{0.05in}

Here, we set the initial added volume of all cells to zero so the
condition in step 1 above is automatically satisfied at $t=0$.  For
our runs, we used 10 cells of initial volume $0.5$ and $t_{\rm max}=T$
is the same as the maximum time for the numerical PDE experiments. We
can also generalize the model to incorporate the mother-daughter
growth coefficient correlation by including a new label $\lambda_i$ to
each cell.

\bibliographystyle{siamplain}
\bibliography{bibliography}
\end{document}